\documentclass[english,reprint, aps,showkeys]{revtex4-1}
\usepackage[T1]{fontenc}
\usepackage[latin9]{inputenc}
\usepackage{amsmath}
\usepackage{amssymb}
\usepackage{graphicx}
\usepackage{hyperref}

\makeatletter

\pdfpageheight\paperheight
\pdfpagewidth\paperwidth

\makeatother

\usepackage{babel}
\begin{document}

\preprint{}

\title{Opinion dynamics model based on cognitive biases}

\author{Pawel Sobkowicz}
\email{pawelsobko@gmail.com}

\affiliation{KEN 94/140, 02-777 Warsaw, Poland}
\begin{abstract}
We present an introduction to a novel model of an individual and group
opinion dynamics, taking into account different ways in which different
sources of information are filtered due to cognitive biases. 
The agent based model, using
Bayesian updating of the individual belief distribution, is based on
the recent psychology work by Dan Kahan. Open
nature of the model allows to study the effects of both static and time-dependent
biases and information processing filters. In particular, the paper compares the effects of two important psychological mechanisms:
the confirmation bias and the politically motivated reasoning. Depending
on the effectiveness of the information filtering (agent bias), the
agents confronted with an objective information source may either
reach a consensus based on the truth, or remain divided despite
the evidence. In general, the model might provide an understanding into the increasingly
polarized modern societies, especially as it allows mixing of different
types of filters: psychological, social, and algorithmic. 
\end{abstract}

\maketitle

\section{Introduction}

The actual processes through which individual people and groups of
people evaluate information and form or change their opinions are
very complex. Psychology offers many descriptions of these processes,
often including multiple pre-conditions and influencing factors. The
assumption that opinions form through a truth-seeking, rational reasoning
is, unfortunately, not true in most cases. The list of the recognized
cognitive biases that influence our mental processes (rational and
emotional) is very long, covering over 175 named entries (\citet{benson2016cognitive}).
The situation becomes even more complex when we try to describe how
the individual opinion changes combine to form dynamical social systems.
In addition to the problems alluded to above, one has to consider
the multiple forms of social interactions: personal (fact to face
and, especially in recent years, those mediated by electronic media)
and public (news, comments, rumours and other modes of information
reaching an individual). These interactions vary with respect to their
informative and emotional content, trust to the source of the information,
its pervasiveness and strength and more. Taking these difficulties
into account, the task of an accurate description of the individual
and group opinion change dynamics appears insurmountable. Yet, the
need to understand how and why our societies (especially democratic
ones) arrive at certain decisions, how and why people change their
beliefs (or why they remain unconvinced in the light of `overwhelming
evidence'), what are the mechanisms driving the increasing polarization
of our societies and how to make people talk to and understand each
other, is so great that despite the challenges, there is intense research
on the topic.

For several years, group opinion change has been a fertile ground
for sociophysics and Agent Based Modelling. The initial works have
used many of the tools and ideas developed to describe magnetic phenomena
and used the analogies between atomic spin states and opinions, magnetic
field and external influences to derive statistical descriptions of
global opinion changes. There are many approaches, for example the
voter model \citep{cox86-1,bennaim96-1,galam02-4,castellano03-1}, 
the Sznajd model \citep{sznajd00-1,stauffer01-2,stauffer02-1,stauffer02-2,slanina04-1,sabatelli03-1,sabatelli03-2,bernardes01-1},
the bounded confidence model \citet{deffuant00-1,deffuant02-1,weisbuch03-1,weisbuch03-2},
the Hegelsmann-Krause model \citep{hegelsmann02-1}, the social impact
model of Nowak-Latané \citep{nowak90-1,nowak96-1} and its further
modifications including the role of leaders \citep{holyst01-1,kacperski00-1,kacperski99-1,sobkowicz10-2},
and many more others. Historically, the initial focus was on the formation
of consensus --- treated as a form of a phase transition ---
but the later works focused on the role of minorities, with special attention
given to the effects of presence of inflexible, extremist individuals. 

The literature on numerical models of opinion dynamics has grown enormously
in the past decade. For a relatively recent reviews we point out \citet{castellano07-1,castellano12-1,galam2012sociophysics}.
While most of the early works were limited to studies of the \textbf{models
themselves} (rather than specific social contexts), showing very interesting
sociophysical results, but only weak, qualitative correspondence to
any actual societies (\citet{sobkowicz09-1}), the recent years have
changed this situation. Availability of large scale datasets, documenting
opinions and interactions between people (derived mainly from the
Internet and social media), has allowed, in principle, to attempt quantitative
descriptions of specific opinion evolution processes.
The number of sociophysical and ABM based works aimed at quantitative description of real societies remains limited. 
For example, in the case of political elections, only a few papers attempt such description (\citet{caruso2005opinion,fortunato07-1,fonseca2015political,palombi2015voting,sobkowicz2016quantitative,galam2016trump}).

Despite the undoubted advances, \textbf{the sociophysical models of the individual
behaviour are still rather crude}. Most of the sociophysical agents and descriptions of their individual behaviour are too simplistic, too much `spin-like',
and thus unable to capture the intricacies of our behaviours. 
This observation applies also to the descriptions of the
interactions between the agents, or, in more general way, to the way
that new information is treated in the process of adjusting currently
held opinions. Most of the Agent Based Models assume relatively simple
forms of such interactions, for example rules which state that if
an agent is surrounded by other agents holding an opinion different
than its own, it would change it opinion to conform to the majority.
As experience with real life situation shows, such `forced' conversion
is rather unlikely among people (in contrast with atomic spins\ldots).
The differences between the model behaviour of spin-persons (\emph{spinsons,}
\citet{nyczka13-1}) and our understanding of real people have forced
the introduction of special classes of agents, behaving in a way
that is different from the rest: conformists, anticonformists, contrarians,
inflexibles, fanatics... Using appropriate mixtures of `normal' and
special agents it has been possible to make the models reproduce more
complex types of social behaviour.

In the author's opinion, such artificial division of the agents into
separate classes with different, fixed internal dynamics, while improving
the models' range of results, is psychologically incorrect. In a specific
situation any person may behave inflexibly or show contrarian behaviour.
For this reason, the author has proposed a model in which opinion
change results from a combination of agent's information and emotional
state, coupled with the informative and emotional content of the message
processed by the agent (which may originate from an interaction with
another agent or from the media). The model, introduced in \citet{sobkowicz12-7,sobkowicz13-2}
has allowed a quantitative description of an Internet discussion forum
\citep{sobkowicz13-3} and even to predict the results of recent elections in
Poland \citep{sobkowicz2016quantitative}. 
The model applies however
only to situations in which the emotional component is very strong,
determining the individual behaviour. 

One of the most active discussions in psychology of belief dynamics
is centred around apparently irrational processing of information:
the operation of biases, heuristic shortcuts and other effects that
stand in contrast with the classical tenets of the rational choice
theory. Such ostensibly irrational behaviours have not received much
attention within the ABM community so far, despite their presence
in many social situations. Important examples are provided by strong
opposition to well documented arguments in cases of climate change,
vaccination, energy policies etc. There are well known differences
in risk perception and reactions, leading to strong polarization almost
beyond capacity to communicate (\citet{tversky74-1,tversky81-1,tversky86-1,opaluch89-1,tversky1990causes,kahneman11-1,sunstein00-1,sunstein02-2,sunstein2006availability,sunstein2016people}).
Our current work has been motivated by the recent studies (\citet{kahan2016politically1,kahan2016politically2}),
which describe in detail the Politically Motivated Reasoning Paradigm
(PMRP). We aim to  create an Agent Based Model using biased information processing and  Bayesian updating. 
Despite the recognized
status of the Bayesian updating in risk assessment and other areas, 
it is rather seldom used by the ABM community.
To mention a few examples, \citet{suen2004self} has considered the effects of information
coarsening (due to the agents' reliance on specialists for the relevant
information) and the tendency to choose the sources which confirm
their pre-existing beliefs; \citet{martins2009bayesian} has studied
the case of continuous opinion model under Bayes rules, looking for
long term evolution of the opinions; \citet{bullock2009partisan}
has studied the conditions in which peoples' beliefs, updated using
Bayesian rules could, in the short term, instead of converging on
a true value, diverge or even polarize. 
\citet{ngampruetikorn2016bias} have analysed the role of confirmation bias in consensus formation in a binary opinion model on a dynamically evolving network.

The flexibility offered by
the Bayesian approach allows much greater complexity of the behaviour
of the individual agents, and as such, offers potentially more relevant
descriptions of social behaviours than the spin-based models. Of course,
these benefits do not come without a price: there are many more
degrees of freedom in the system, and therefore many more unknowns
in properly setting up the ABM simulations. Still, the importance
of  social phenomena observed around the world, in particular various
forms and effects of polarization, suggests the need for a deeper
understanding of the underlying mechanisms, and makes the effort worthwhile. 

\subsection{Confirmation bias vs. PMRP}

One of the best recognized biases in information processing is 
\textit{confirmation bias}, defined by the Wikipedia as `a tendency to search
for, interpret, favour, and recall information in a way that confirms
one's pre-existing beliefs or hypotheses'. Such definition stresses
that the operation of the confirmation bias may be on various levels:
selecting and preferring the information sources, giving different
weight to different sources and internal mechanisms (such as memory
preferences is storing/recall of information). When people communicate,
the individual confirmation bias effects may be combined in a way
that creates group effects such as echo chambers. As a result, even
when faced with true information, and agent (or a group of agents)
may form or maintain a false opinion due to the confirmation bias.

The \emph{motivated reasoning } paradigm considers the ways in which
goals, needs and desires influence the information processing (\citet{jost2013hot}).
These goals may be related to the individual needs, but also to group
or global ones, for example the goal of \textit{achieving or maintaining the
person's position within a social group}. In such a case, the motivated
reasoning may bias the information processing by substituting the
goal of truth-seeking by the person's desires to affirm the affiliation
with the chosen in-group. Seen through the lens of these desires,
the apparently irrational choices (such as disbelief in well documented
evidence and belief in unproven claims) become rational again. We
believe and act accordingly in a way that is congruent with the perceived
beliefs and actions of our preferred social group. As the goal of
the person is shifted from truth seeking to strengthening of the position
within the social group, the disregard for the truth becomes rational.
Especially, when the consequences of a rejection from the group are
more immediate and important than the results of `erroneous' perception
of the world. \citet{kahan2016politically1,kahan2016politically2}
has provided a very attractive Bayesian framework, allowing not only
to describe the role of various forms of cognitive biases, but also
the empirical evidence of the differing predictions of the different
heuristics, such as confirmation bias or political predispositions.
Experiments with manipulated `evidence', described by Kahan, are very
interesting.

While both mechanisms introduced above lead away from the truth-seeking
behaviour, their predictions might differ, especially with respect
to new information. While the confirmation bias favours evidence in agreement with already held views (priors),
the politically motivated reasoning selects and favours information
congruent with person's political identity (defined by the in-group characteristics).
The confirmation bias depends on internal agent states, while PMR involves perception of external characteristics. 

The vision of information processing, comparing the two forms of bias,
described by Kahan is simple enough to become a framework of an
ABM. As we shall argue, the Bayesian filtering approach is very flexible
and may be applied to a variety of situations, contexts and types
of processing bias. \textbf{Our present goal is to describe such framework
and to provide simple examples of the types of information processing leading
to consensus or polarization.} The latter case is of special importance,
as the current political situation in many democratic countries seems
to be irrevocably polarized, with large social sections unable to
find common ground on many extremely important issues. Our hope is
to find, with the use of the model, any suggestions for the processes
that may reverse this polarization and enable communication across
the current divisions.

\section{Individual information processing model}

\subsection{Overview of the model}

\begin{figure*}
\includegraphics[width=0.8\textwidth]{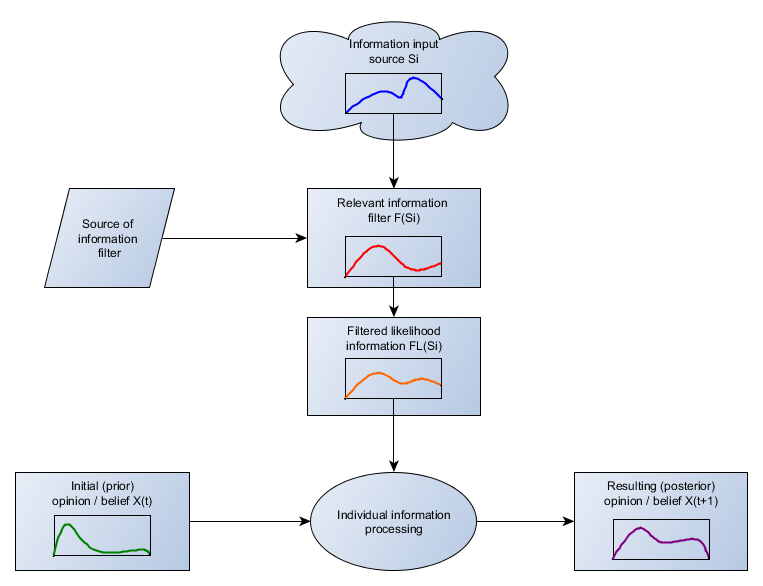}

\caption{Basic model of information processing. An agent holds a prior belief
about an issue, described by a distribution $X(\theta,t)$. We assume
a simple, one-dimensional `opinion parameter' $\theta$ ranging from
-1 to 1.The information on the issue, coming from the source $S_{i}$
has a distribution $S_{i}(\theta)$.This information is filtered by
a function $F(S_{i})$, specific to the information source. The form
of the filtering function may vary, depending on the focus of the
model. For example, if we assume fully rational, truth-seeking agents,
the filter would be centred around the `true' value of the parameter
$\theta$. On the other hand, in the case of the model based on cognitive
bias, the filter function would be simply related to the prior beliefs
of the agent. In the case of PMRP, the filter is related to the distribution
of beliefs held / approved by the agent's in-group, or, more precisely,
to the agent's perception of such distribution. Combining the information
input with the filter function yields the filtered likelihood information
$FL(S_{i})$. Bayesian update of the agent's belief $X(t)$ via $FL(S_{i})$
leads to the changed, posterior distribution of beliefs $X(\theta,t+1)$.
\label{fig:singlesource}}
\end{figure*}

The current work aims at a general, flexible \textbf{model of the individual
opinion dynamics}. We base our concepts on the Bayesian framework.
Figure \ref{fig:singlesource} presents the basic process flow, modelled
after \citet{kahan2016politically1}. For simplicity, we shall assume
that the belief which we will be modelling may be described as a single,
continuous variable $\theta$, ranging from -1 to +1 (providing
a natural space for opinion polarization). The agent holds some
belief on the issue, described at time $t$ by a distribution $X(\theta,t)$.
For example, if the agent is absolutely sure that the `right' value
of $\theta$ is $\theta_{0}$, then the distribution would take the
form of Dirac delta function centred at $\theta_{0}$. Less `certain'
agents may have a different form of $X(\theta,t)$. This distribution
is taken as a prior for a Bayesian update, leading to the opinion at $t+1$. 
In the simplest case, the
Bayesien likelihood factor would be provided by the new information
input $S_{i}(\theta)$. 
Here the index $i$ corresponds to various possible information sources.
Kahan has proposed that instead of this direct
update mechanism (prior opinion+information$\rightarrow$posterior opinion), the incoming information is filtered by the cognitive
biases or predispositions of the agent. The filtering function $F(S_{i}$)
transforms the `raw' information input into the filtered likelihood
$FL(S_{i},\theta)$, so that the posterior belief distribution $X(\theta,t+1)$
is obtained by combining the prior opinion distribution $X(\theta,t)$ with the likelihood filter  $FL(S_{i},\theta)$. 

\begin{figure*}
\includegraphics[width=1\textwidth]{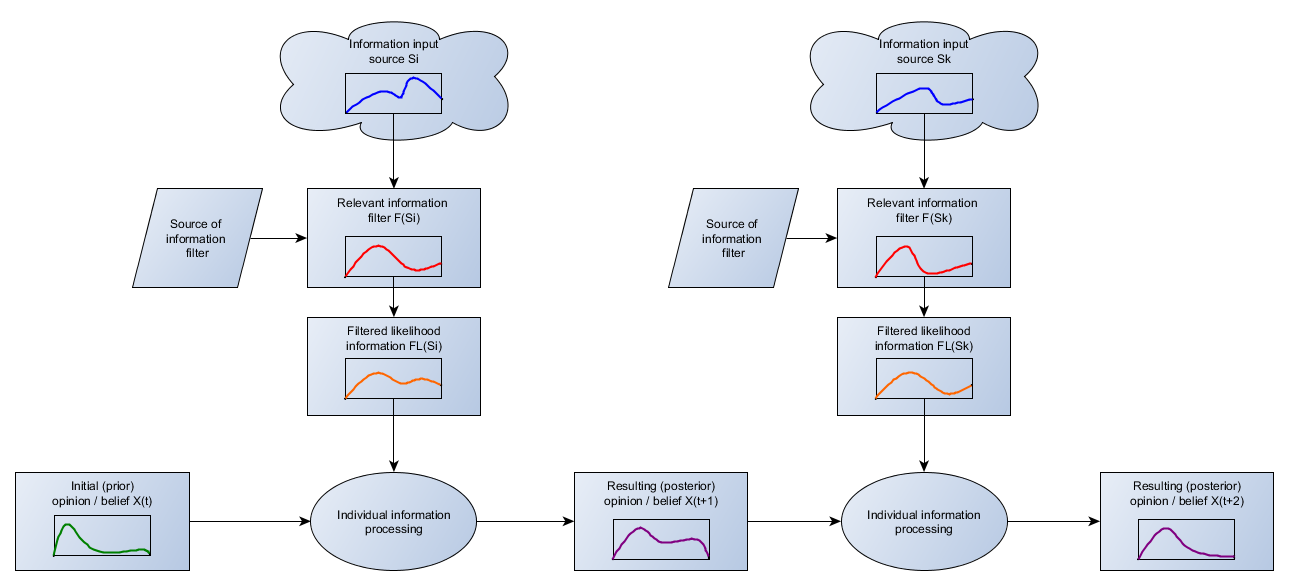}

\caption{Sequential model of information processing when multiple sources are
present. As before, an agent holds a prior belief about an issue,
described by a distribution $X(\theta,t)$. We assume a simple, one-dimensional
`opinion parameter' $\theta$ ranging from -1 to 1.The information
on the issue, coming from the source $S_{i}$ has a distribution $S_{i}(\theta)$.
A different distribution $S_{k}(\theta)$ may come from another source
$S_{k}$. There filtering functions $F(S_{i})$ and $F(S_{k})$ may
differ, and as a consequence, the likelihood functions $FL(S_{i})$
and $FL(S_{k})$ would also differ. Bayesian update of the agent's
belief $X(t)$ via $FL(S_{i})$ and $FL(S_{k})$ is applied sequentially,
leading to the changed, posterior distribution of beliefs $X(\theta,t+1)$
and $X(\theta,t+2)$. In the case of many sources, their relative
importance may be described by the number of times they are present
in the chain of evaluations. \label{fig:sequential}}
\end{figure*}

\begin{figure*}
\includegraphics[width=0.8\textwidth]{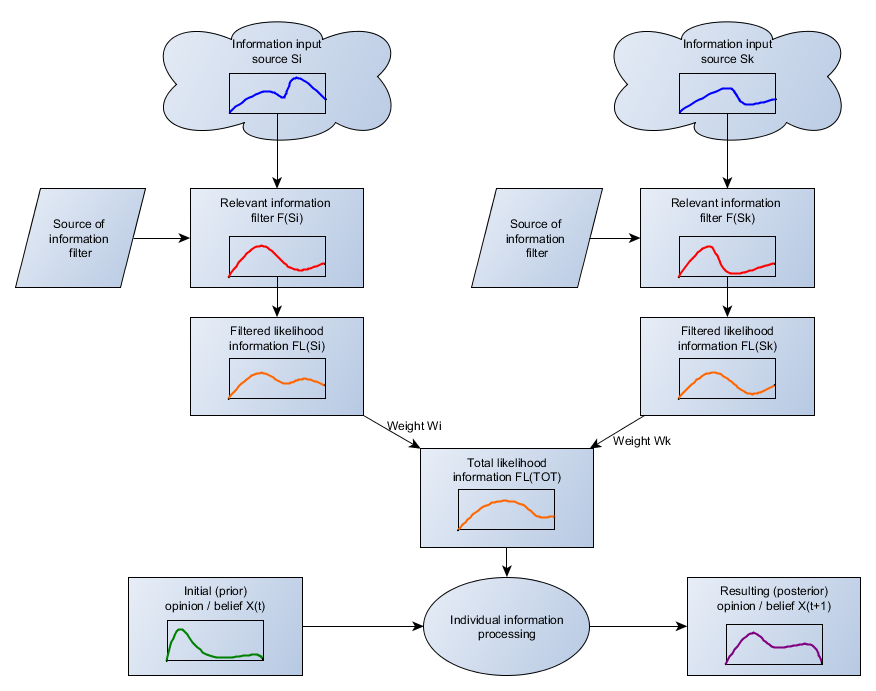}

\caption{Integrative model of information processing when multiple sources
are present. As before, an agent holds a prior belief about an issue,
described by a distribution $X(\theta,t)$. The information on the
issue, coming from the sources $S_{i}$ and $S_{j}$ has a distributions
$S_{i}(\theta)$ and $S_{j}(\theta)$. There filtering functions $F(S_{i})$
and $F(S_{k})$ may be different. As a consequence, so would the likelihood
functions $FL(S_{i})$ and $FL(S_{k})$. Instead of a sequential application,
the Bayesian update of the agent's belief $X(t)$ via a single application
of a weighted combination of $FL(S_{i})$ and $FL(S_{k})$: $FL(TOT)=W_{i}FL(S_{i})+W_{k}FL(S_{k})$,
leading to the changed, posterior distribution of beliefs $X(\theta,t+1)$.
The weights determine the relative importance of the different information
sources in a single update. \label{fig:integrative}}
\end{figure*}

It is important to note that different sources of information may
be filtered in different ways. Trust in the source, cognitive difficulty
of processing the information, its emotional context, the agent's
dominant goals -- they all may influence the `shape' of
the filter. Moreover, we have to consider the ways that the information
pieces from various sources are treated. Two simple versions are shown
in Figures \ref{fig:sequential} and \ref{fig:integrative}. The first
treats each source separately and in a sequential order. Such approach
may be sensible in cases where new information arrives in well separated,
time ordered units, e.g. daily newspaper editions or TV news programs.
The second approach treats the sources in an integrative way:
it accumulates the filtered likelihoods (each of which contains the
information and its specific filter), with some weights, into a single
total likelihood function. Such approach may be better when various
sources of information coexist at the same moment, e.g. when a group
of people discusses the TV news. The weights associated with the sources
could be different for each information processing event, depending
on the relative importance and strength of the sources and other circumstances.
Both approaches can be used (and combined) in the case of advanced
models of specific systems.

\subsection{Information sources}

The information that influences the beliefs of people comes from multiple
types of sources. There are, of course, \textbf{personal experiences},
which may provide high impact information about specific facts and
events, and, with the application of some cognitive processes, about
trends, estimates, diversity and prognoses. The direct experiences
may be thought of as direct and therefore trustworthy, but in many
cases we rely on memories, which may provide false information. Some
other cognitive biases are also relevant for the personal observations:
we may fall for certain illusions, disregard a part of experience
and put emphasis on other parts, even to a degree of actually inventing
events that did not take place. 

The second source of the information is related to the group of 
people with whom a person identifies (the in-group). These inputs
may come from \textbf{in-group information exchanges}, either in person
or via electronic or traditional communication media. The latter has
become increasingly important during the past decade, especially among
the younger population. In addition to the interactions with specific
individuals in the in-group, the in-group may influence agents beliefs
via cumulative indicators. These would include the \textbf{official
or semi-official statements of the group's views on specific issues},
but also the \textbf{unofficial and media information about the group
norms, average opinions and trends}. The latter are especially interesting,
as they may come both from within the group and from outside. 
In such case the information about the in-group views and norms may be manipulated and distorted.

The last group of the sources is related to any source outside the
in-group. This may include the \textbf{interactions with people outside
one's own self-identification group} and the \textbf{media} perceived
as not associated with the in-group. In  case of the media the information
is \textbf{prepared} by someone, which includes both the selection and presentation of the information. 

The information which we use to fortify or
to change our beliefs may be manipulated `at source'. In personal
interactions with other people we may get the wrong impressions because
people due to many forms of dishonesty or distortion. Traditional
sources of news are also subject to misrepresentation. The ideal of
the fair and balanced journalism \textendash{} giving comparable attention
to all contradicting views \textendash{} may also, at times, be considered
manipulative, especially when it results in undue attention and coverage
given to a tiny minority of views. A example of negative consequences
of such `balanced' reporting may be provided by the case of the anti-vaccination
movement (\citet{betsch2013debunking,nelson10-1,tafuri2013addressing,wolfe02-2}). 

In reality, however, much more frequent are the manipulations due
to \textbf{unbalanced reporting}. The polarization of both the traditional
channels (newspapers, radio, TV) and the Internet sources (WEB versions
of the traditional channels and independent WEB pages, blogs, Facebook
pages and tweets) is a well known phenomenon (\citet{adamic05-1,stroud10-1,campante10-1,lawrence10-1,jerit2012partisan,prior2013media,wojcieszak2016partisan}).
Many people rely on a limited number of information sources, the spectrum
of the information reaching him/her could be heavily distorted. Theis
selective attention/selective exposure may lead to the echo-chamber
phenomenon, where a person sees and hears only the information supporting
the `right' beliefs.

The US presidential election in 2016, with its increasing role of
social media as information sources, brought our attention to yet another
form of `at source' information manipulation: \textbf{fake news}.
The relative ease to create false information, in some cases supported
by manipulated images, voice and video recordings, to post it online
and to create a web of self-supporting links allows the perpetrator
to spread such news. The trust associated with social networks (for
example Facebook or twitter links) makes spreading of such information
faster -- especially if the fake news are designed to pass
through the most common information filters.

As examples, we propose  three specific forms of the
distributions of the source information $S(\theta)$, suitable for the ABM approach. The first, $S_{T}(\theta)$,
corresponds to results of social efforts to describe the phenomenon
as \textbf{accurately and objectively} as possible. Let's assume that
for the topic in question (where the beliefs are described by the
parameter $\theta$) there is some specific value that corresponds
to objectively discoverable optimum, $\theta_{T}$. This may correspond
to an exact description of a situation, or an universally optimal
solution to a problem or any other situation in which, through rational,
communicative processes, it is possible to arrive at a `true' value
of $\theta_{T}$. This would mean, that eventually, all beliefs other
than this value should be labelled `erroneous'. We assume that $S_{T}(\theta)$
takes a form of a Gaussian distribution centred around $\theta_{T}$. 
In the following simulations we shall use $S_{T}(\theta)$ as the source.

The second possible form of a popular information distribution, a `flat' one 
$S_{F}(\theta)$, assumes that all possible values of $\theta$ are
represented equally in the information stream (an extreme version
of the `\textbf{fair and balanced}' news). 

\begin{figure}
\includegraphics[width=1\columnwidth]{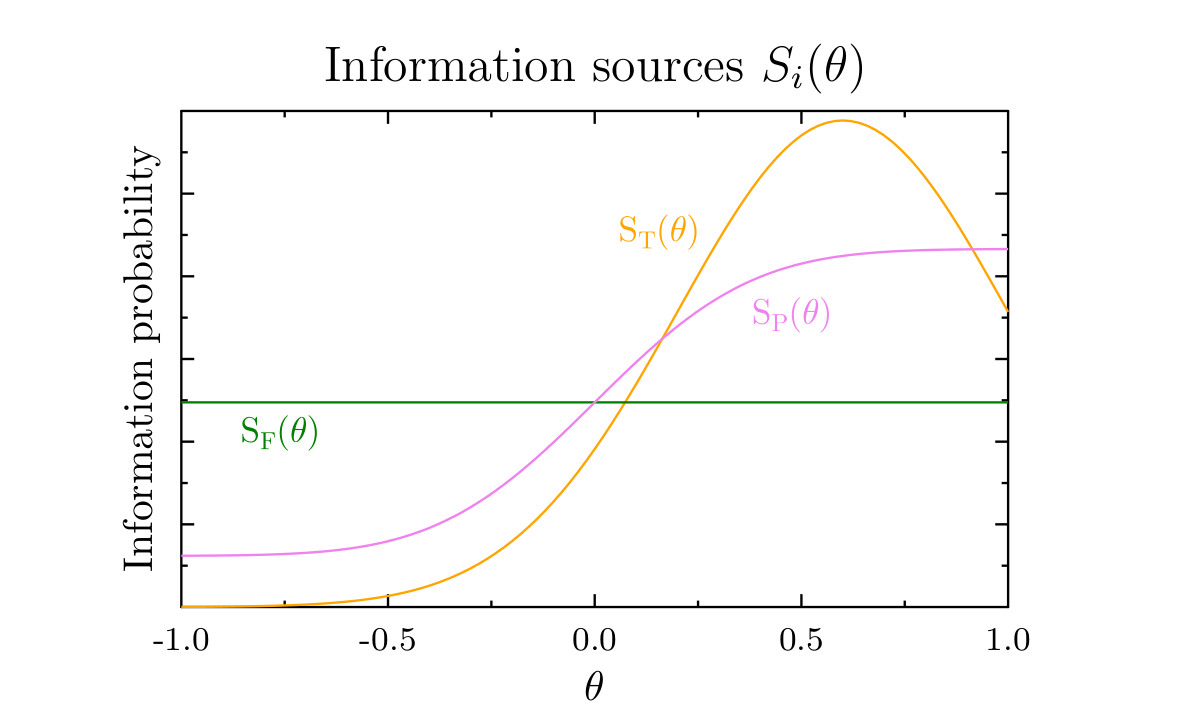}

\caption{Graphical representation of examples of information sources. $S_{F}(\theta)$
\textendash{} the flat distribution representing an extreme form of `balanced'
news. $S_{T}(\theta)$ \textendash{} an example of a distribution
focusing on a `true' value accepted by the whole community. In this
example, the true value is set at $\theta=0.6$ and the information
distribution has a rather broad shape. Lastly, $S_{P}(\theta)$ represents
partisan bias, in this case favouring positive $\theta$ values at 6:1
ratio.\label{fig:sources}}
\end{figure}

The third form, $S_{P}(\theta)$,
is designed to represent \textbf{partisan} bias in the news stream,
taking a form of a sigmoidal function, favouring one of the alternatives
(for example $\theta>0$). The three forms are shown in Figure \ref{fig:sources}.

\subsection{Types of information filters}

The way in which information received from various sources is evaluated
and used to form new beliefs, depends not only on the sources, but
also on the goals of a person. These goals may allow us to construct
rules that would create and update the information filters. In some
cases they would be independent of the characteristics of the
person, in other cases they would depend on them, which would make
the process of belief modification self-referential. Below is a partial list
of the filter types that could be used in our agent based modelling.
The filters are distinguished by their origin (internal to the person
or external), dependence on some objectively measurable characteristics,
possibility of an orchestrated manipulation and, finally, normative
value. 
\begin{itemize}
\item \textbf{Truth seeking filter}. It corresponds, on an individual level,
to the objective source of information. The truth seeking filter could
take a form of a distribution localized around $\theta_{T}$, so that
the eventual, repeated application of the information processing would
lead the agents to converge their beliefs on $\theta_{T}$. An example
of such filter would be a narrow Gaussian distribution centred on
$\theta_{T}$. This type of the filter is at the core of the `rational
discourse' and `objective reality' assumptions, and while important
from the philosophical and moral standpoints it seems to be an exception
rather than a rule in social life. Because the value of $\theta_{T}$
is independent of the agents, we categorize the truth seeking filter
into the `external' category. And because the discovery of the value
is assumed to rely on well established processes (for example depending
on scientific methods), we also assume that the truth seeking filter
in its pure form is not liable to manipulation. Applied to the flat
(balanced) information $S_{F}(\theta)$ the truth seeking filter would
create a filtered information resembling the objective, truth-related
source $S_{T}(\theta)$.
\item \textbf{Confirmation bias filter}. The tendency to give more weight
to information supporting a person's current views and to disregard
sources disconfirming these views is well known in psychology. Such
filter is relatively easy to be introduced into the ABM framework:
the filtering function $F_{CB}(S(\theta))$ could be defined in terms of
the current belief $X_{j}(\theta,t)$, either directly or slightly modified,
for example via some form of broadening or 
narrowing of tolerance range for differing
beliefs. A very widely known example of the use of confirmation bias in AMB environment is the bounded confidence model
(\citet{deffuant02-1,hegelsmann02-1,weisbuch03-1}). In the model an agent
interacts only with other agents who have an opinion sufficiently
close to its own, disregarding agents with opinions separated by more
than a certain threshold $\epsilon$. The confirmation bias filter
belongs to the internal category, and as such, it can not be manipulated
directly by outside sources. A possibility exists, however, that the
manipulation would act on the importance or tolerance of the filter
in evaluating certain sources of information.
\item \textbf{Memory priming/availability filter}. It is another example
of an internal filter, which, however, is much more easily manipulated
than the confirmation bias filter. This is because the confirmation bias compares the
new information with currently held beliefs, which may be quite deeply
ingrained, especially if they depend on moral foundations (\citet{jost03-2,haidt2007new,haidt12-1,haidt2012righteous}).
In contrast, the availability filter acts via additional attention given to facts that are quickly
accessible to our minds. Thanks to various forms of priming, its effects may be
effectively stimulated and steered by outside influence: our peers or the media
(\citet{tversky74-1,tversky96-1,sunstein2006availability}). In terms
of an ABM approach, such filter could be approximated, for example,
by the shape of the previously encountered information source.
\item \textbf{Politically Motivated Reasoning (PMR) filter}. The notion
of the PMR filter advocated by \citet{kahan2016politically1}, is
based on an assumption of a perfectly rational behaviour \textendash{}
but with a re-defined personal goals. Instead of the focus on the
exact description driving the truth-seeking filter, the rationality
of a person's actions is judged by their usefulness for the goal of
preserving or improving the position within a specific social group
(the \textit{in-group}). 
In such case, the dominant processes would be those which
facilitate alignment with the in-group acceptance criteria, which
often include expression of specific beliefs. Thus the PMR filter
would be based on the \textbf{perceived} in-group opinions. 
As such, the PMR filter is an example of an external filter,
that is, in itself, based on some information source, rather than on the internal characteristics of the agent. 
For example it
could be a Gaussian distribution centred at the average belief of
the in-group. It is
worth noting that in some cases \textbf{ it can be manipulated}. 
The range of such manipulation depends on a specific
social context. Even in the cases when the knowledge
about the in-group beliefs comes from direct interactions between
the members of the in-group, some external social pressures might
limit the expression of these beliefs. For example, political correctness
might prevent overt expressions of some opinions, leading to a departure
of the perceived average value from the `true' average of internally
held, but not expressed, beliefs. Another possibility of manipulation
is when the information about a large in-group (such as political
party support base) is mostly available via some external media: press,
TV, social networks... The medium may withhold some information, enhance
some other and in such way distort the perceived in-group opinions
and thus manipulate the agent's PMR filter.
\item \textbf{Simplicity/attention limit filter}. This is an internal
filter, related to the culturally and technologically driven change
in the way external information is processed. Due to the information
deluge, there is an increasing dominance of short forms of communication,
especially in the Internet based media: WEB pages, Internet discussions,
social media (\citet{djamasbi2011online,djamasbi2016text}). 
The simplification
(or oversimplification) of important issues, necessary to fit them to the
short communication modes, may act against beliefs that are not disposed
to such simplification. This part of the filter acts at the creation
side of the information flow. Decreasing attention span and capacity
to process longer, argumentative texts act as another form of filter,
this time at the reception end of the flow. There are numerous forms
of psychological bias related to and leading to such filtering, from
venerable and accepted heuristics (like Occam's razor), through the
law of triviality and bike-shed syndrome (\citet{parkinson58-1}), to a total disregard for
too complex viewpoints (\citet{qiu2017lack}). Together, these tendencies
can create a filter favouring the information that is easily expressed
in a short, catchy, memorizable form. There is no simple universal
form of the filter for the ABM approach, because in different contexts
different beliefs might be easier to be expressed in the most simple
way.
\item \textbf{Emotional filter}. Some topics, contexts and communication
forms may depend, in their processing, on the affective or emotional
content. This may create a processing filter, for example one that
favours extreme views, as they are typically more emotional than the
consensus oriented, middle-of-the-road ones. Emotionally loaded information
elicits stronger response and longer lasting effects (\citet{hatfield1993emotional,haidt01-1,barsade02-1,allen05-1,clore07-1,berger2010social,nielek10-1,sobkowicz2010dynamics,chmiel11-1,chmiel11-2,tagar11-1,thagard11-1,sobkowicz12-7,bosse2013modelling}).
The specific form of the filter depends on the mapping of the belief
range and the associated emotional values. Furthermore, the emotional
filter may depend on the current agent belief function, e.g. anger
directed at information contrary to the currently held beliefs,
or at a person who acts as the source of the information.
\item \textbf{Algorithmic filters}. An increasing part of the information
reaching us comes from the Internet services such as our own social
media accounts, personalized search profiles etc. The service providers
organize and filter the content that reaches us  often without
our knowledge that any filter exists; and even more often without
the knowledge how it works. These external algorithmic filters, shaping our perception,
not only skew the opinions but, more importantly, they often limit
the range of topics we are aware of and the opinions related
to them (\citet{pariser2011filter,albanie2017unknowable}). In some
cases the effect of an algorithmic filter is similar to the internal
confirmation bias (e.g. the search engine prioritizes the results
based on the already recognized preferences of the user). In other
cases, the machine filter may deliberately steer the user away from
certain information, based on decisions unrelated to the particular
user, fulfilling the goals of some other party. 
\end{itemize}
\begin{figure*}
\includegraphics[width=0.8\textwidth]{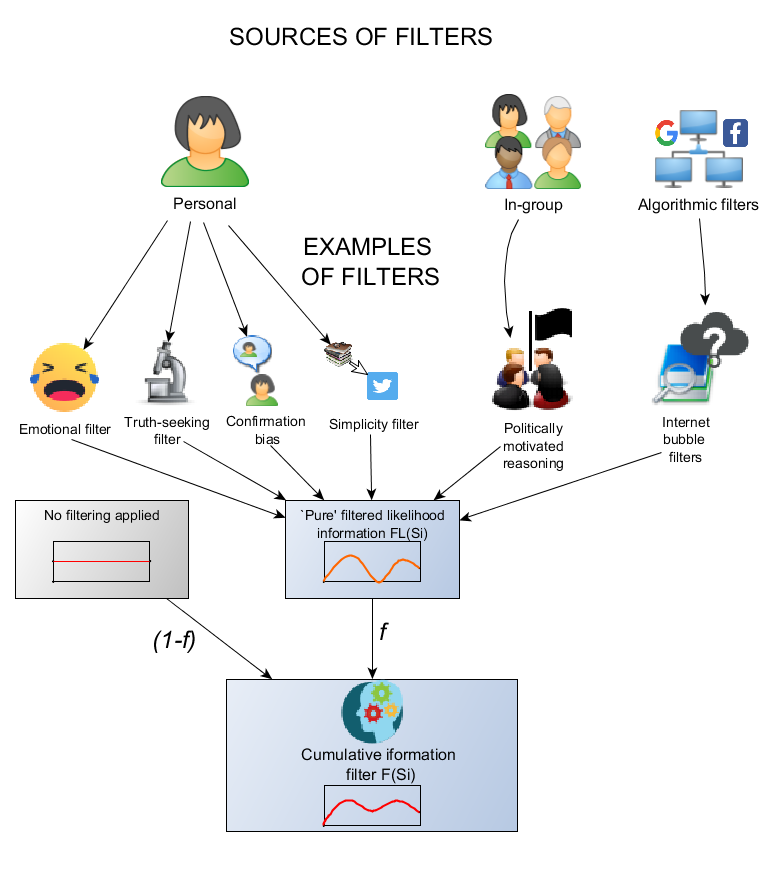}
\caption{Graphical representation of selected filtering types and their sources.
The mix of various forms of filtering may depend on the source of
the information being evaluated, for example in certain situations
the personal, truth-seeking filter might be dominant, while in other
situations the focus on the in-group acceptance would favour the PMR
filter. When media or the Internet are the source of the information,
the personal filters may be modified by the effects of personalized
filtering by the search/presentation algorithms of the content providers
or intentional modifications in the news industry. To allow the possibility
of imperfect application of the filters, in the final stage the `pure'
filtered likelihood may be combined with non-specific, uniform function
$U$, via the filter effectiveness factor $f$. The resulting function
would have the form of $fF+(1-f)U$.\label{fig:filters}}
\end{figure*}

\subsection{The filtering process }

An interesting question, important for the practical ABM implementation,
is: how should the various filters be applied to obtain the relevant
filtering function $F(S_{i},\theta)$? A sequential application of
filters focuses on the parts of the information that are minimized
or deleted by each filter. In contrast, parallel application focuses
on the information that is allowed by each of the filters. In reality
some of the filters are applied sequentially, that is a person considers
only the information that conforms to all of these filters (e.g. it
must be highly emotional \textbf{and} in agreement with person's views).
In other cases, some filters may be added, for example a person may
accept the information that confirms his/her views (the confirmation
bias) \textbf{or} the information that agrees with he perceived views
of the in-group. As a result, the overall shape of the filter function
may become quite complex. 
Moreover, we should remember that even if we treat some external filters
as relatively stable, the ones associated with person's own views
or with the in-group compliance may evolve in time.

\begin{figure}
\includegraphics[width=1\columnwidth]{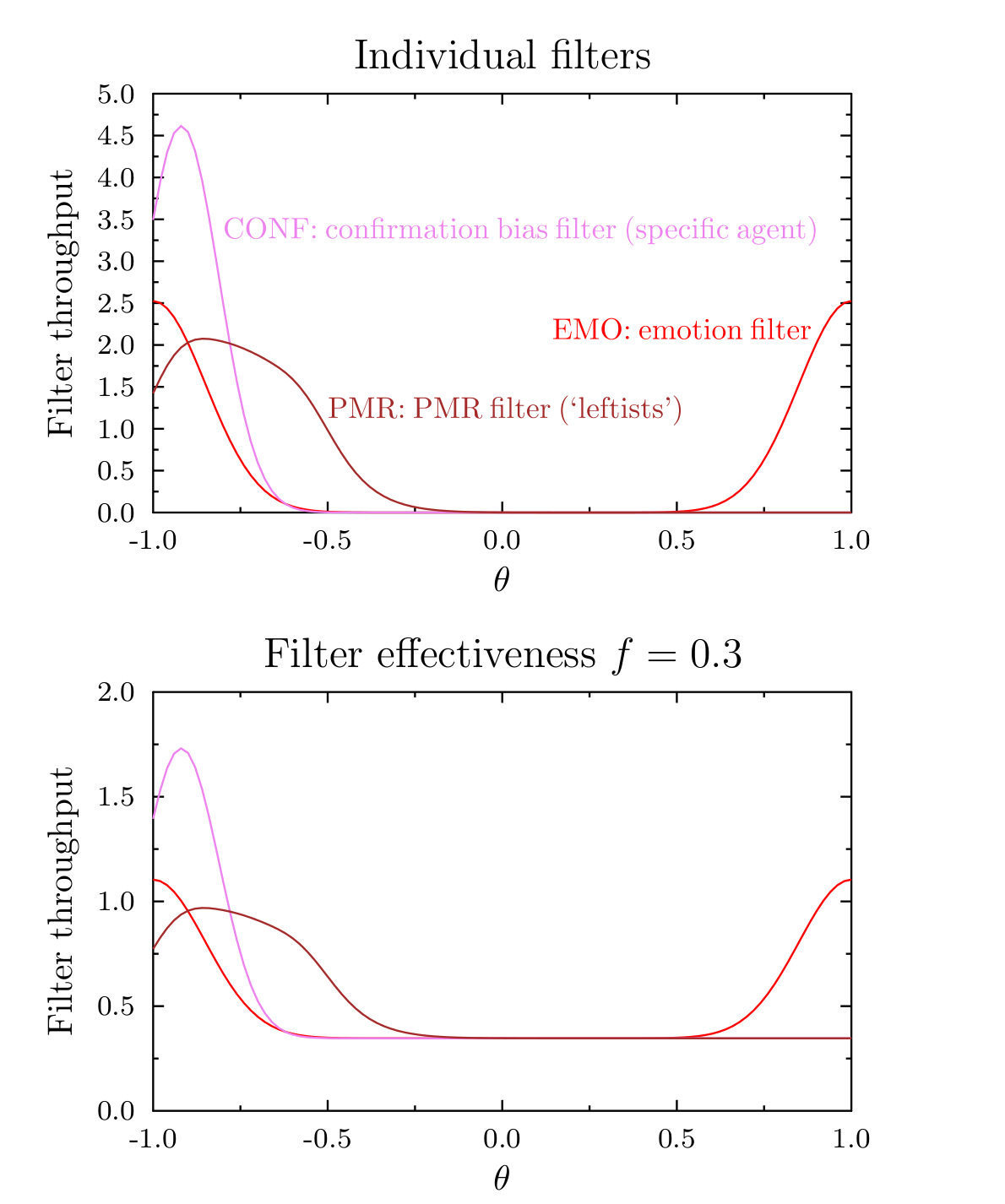}

\caption{Examples of filtering mechanisms. Top
panel:  three `pure' filters (with effectiveness $f=1$). CONF (violet line):
confirmation bias taken as an example of an individual agent's belief,
in this case a rather extreme leftist. PMR (brown line): Politically
Motivated Reasoning filter, is calculated as and average belief of
the agent's in-group (the leftists, in this case). EMO (red line):
an example of a simple emotional filter, favouring extreme views. Bottom
panel: effects of a an imperfect filtering, in which the effectiveness
factor $f$ is assumed to be equal to 0.3 In such case, the resulting
filter function is given by $fF+(1-f)U$, where $F$ is the `pure'
filter and $U$ is a uniform function.\label{fig:filters-1}}
\end{figure}

The Bayesian-like form of the filtering process, multiplying the incoming
information $S_{i}(\theta)$ by the filter function $F(\theta)$
is very efficient: a single process may decisively change the shape
of the information distribution. For this reason we introduce here
a process control parameter, the filtering efficiency $f$. Its role
is to determine the relative strength of the influence of the specific
filtering function on the incoming information. In particular, the
\textbf{effective filter function} is assumed to take the form $fF(\theta)+(1-f)U$,
where $U$ is a uniform function. 

\subsection{Information processing and memory effects\label{subsec:Information-processing} }

The  Bayesian processing of information
may lead to very quick and dramatic revisions of the individual beliefs.
Some shapes of the likelihood distributions (especially those with
a narrow maximum) transform prior belief distributions into completely
different posterior. Yet, with some exceptions, our interactions with
other people or with the media sources are rarely so transformative.
\citet{martins2009bayesian} has proposed a modification of the original
Bayesian rules, relaxing the transformation speed. He has proposed
that only a fraction $p$ of encounters with the information sources
leads to informative processing, characterized by the likelihood
function $FL(S_{i})$. In the remaining $1-p$ cases, Martins has
considered that the source is treated as uninformative (characterized
by a uniform distribution), and the resulting information source is
a mixture of the two, in a way similar to our treatment of the filtering efficiency.

\begin{figure}
\includegraphics[width=1\columnwidth]{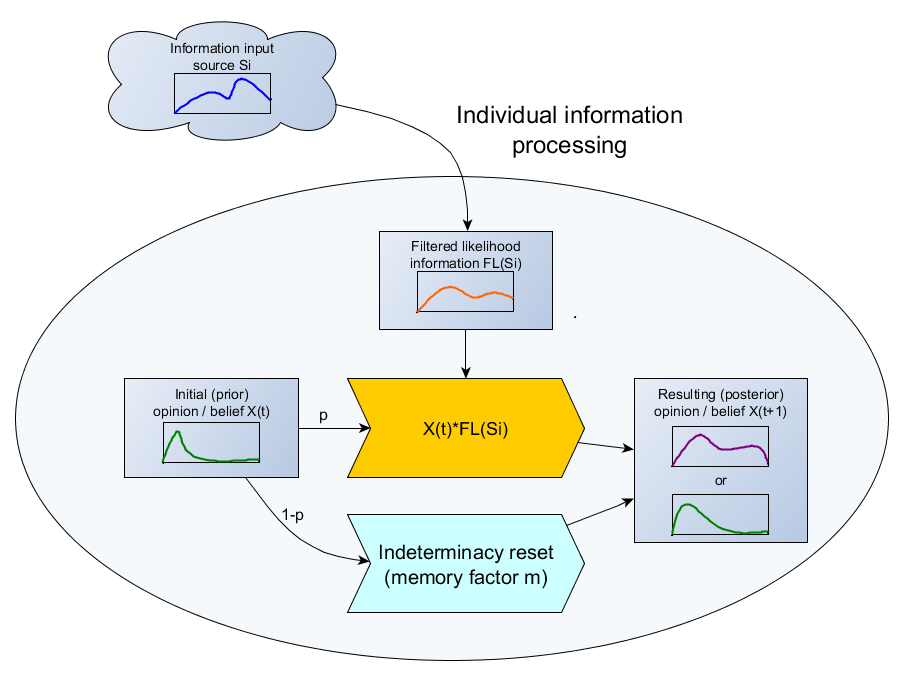}

\caption{Details of the information processing. The likelihood function $FL(S_{i}(\theta)$,
derived from the information source $S_{i}$ is applied only in case
of `informative' encounter, with probability $p$. 
In such case $X_j(\theta,t+1)=X_j(\theta,t) FL(S_{i}(\theta)$
In the remaining
cases, the encounter is ignored. Without processing the new information
the agent's belief function may remain unchanged, or it may become
somewhat relaxed, by addition of a Gaussian function $N(\langle\theta\rangle_{j}(t),\sigma_{0j})$,
centred at the \textbf{current} average $\theta$ for the agent,
and characterized by the standard deviation equal to the \textbf{starting
} value $\sigma_{0j}$. Depending on the value of the memory parameter
$m$ the posterior belief of the agent becomes then $X_j(\theta,t+1)=mX_j(\theta,t)+(1-m)N(\langle\theta\rangle_{j}(t),\sigma_{0j})$.\label{fig:information}}
\end{figure}

In our approach, in the remaining $1-p$ cases, the encounter is \textbf{ignored
and the information is not processed}. The simplest approach to describe
such situation would be to leave the belief distribution unchanged,
$X_j(\theta,t)=X_j(\theta,t+1)$, which may be treated as the case of
the agent's \textbf{perfect memory}. However, as we shall show in the next section, repeated application
the Bayesian updates leads to a narrowing of the agents' belief distributions.
Eventually, the individual beliefs would become more and more focused,
which influences the whole system dynamics. For this reason we will
introduce an \textbf{imperfect memory mechanism} that restores some
level of an \textbf{individual belief indeterminacy}, in which the
agent reverts partially to its intrinsic value of the standard deviation
of the $X(\theta,t)$ distribution. This is described as follows:
in the case of ignoring the information event (probability $1-p$)
the agent's belief distribution does not remain unchanged but becomes

\begin{equation}
X_j(\theta,t+1)=mX_j(\theta,t)+(1-m)\textrm{N}(\langle\theta\rangle_{j}(t),\sigma_{0j}),
\end{equation}
where the memory fidelity parameter $0\leq m\leq1$ describes the
ratio of preserving the current distribution intact, and $\textrm{N}(\langle\theta\rangle_{j}(t),\sigma_{0j}(t))$
is a Gaussian distribution centred at the current average belief
of the agent $\langle\theta\rangle_{j}(t)$, but characterized by a
fixed standard deviation $\sigma_{0j}$, characteristic for each agent.
Thus for the perfect memory ($m=1$) we recover the unchanged distribution
condition and for $m=0$, an agent `left to itself' preserves the current
average value of the belief, but resets the indeterminacy of its beliefs
to $\sigma_{0j}$. The information processing is graphically presented
in Figure \ref{fig:information}.

\begin{figure}
\includegraphics[width=1\columnwidth]{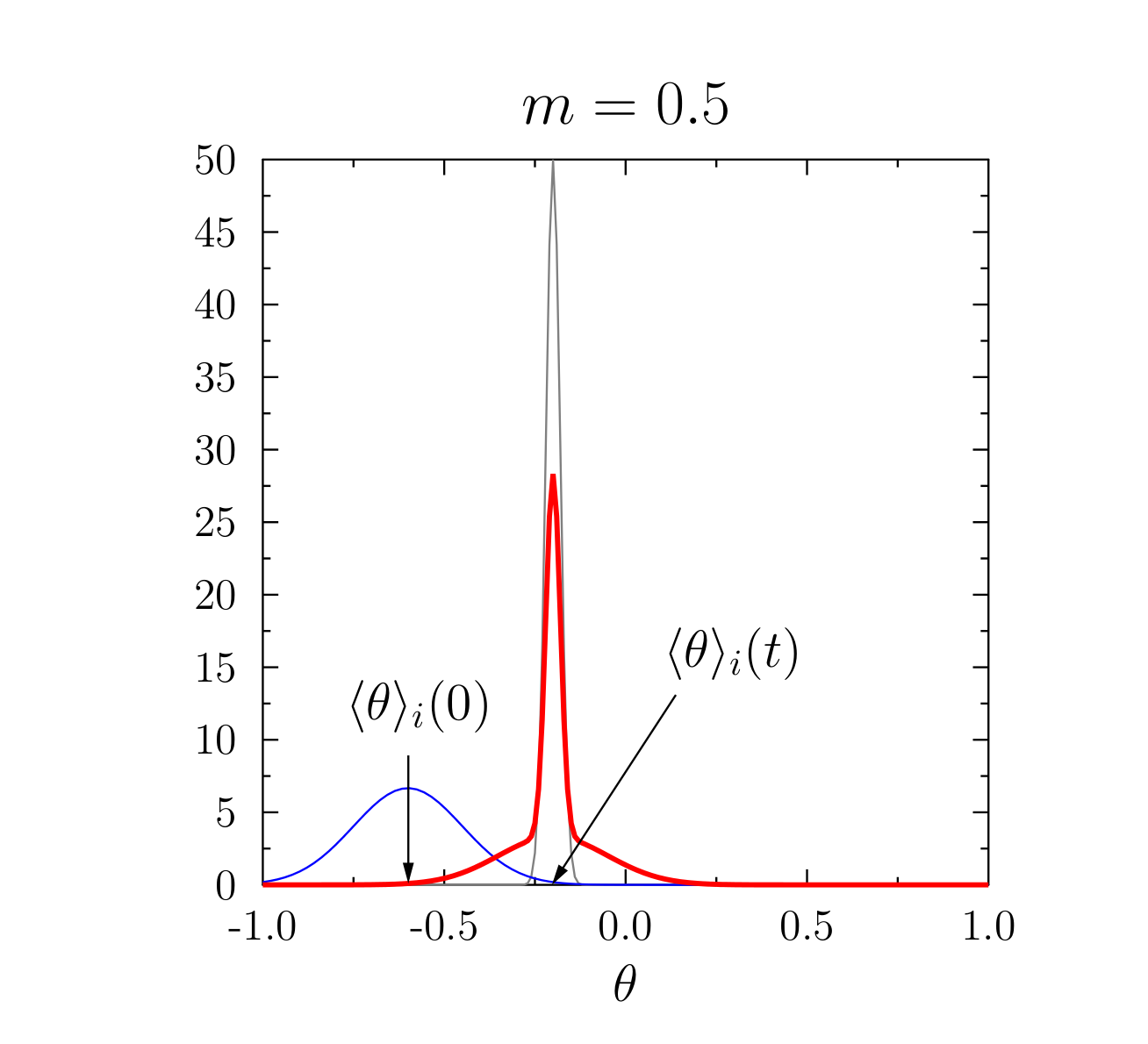}

\caption{Example of the indeterminacy reset (memory factor) of an individual
agent $i$ belief distribution. The current (already quite narrow)
belief distribution (thin black line), centred at $\langle\theta\rangle_{i}(t)$
is mixed with a normal distribution centred at the same value, but
with the standard deviation $\sigma_{0j}$ characteristic for the
agent (blue line shows the original distribution, which is centred
at $\langle\theta\rangle_{j}(0)$). The resulting distribution (red
line) exhibits the central peak but also some `tails' that allow the
agent to accept beliefs further from the centre of its belief function.
\label{fig:memoryeffect} }
\end{figure}

The results of resetting the indeterminacy of an individual agent
belief distribution is shown in Figure \ref{fig:memoryeffect}. While
the `broadening' admixture may seem comparatively small (at least
for the depicted $m=0.5$ value), it plays an important role in shaping
the evolution of the beliefs of the agents under the influence of
information sources. The origin of such reset of the indeterminacy
may be explained by numerous encounters with a range of beliefs, other
than the main source considered in the simulations, which are too
weak to significantly shift the agent's average opinion, but introduce
some degree of uncertainty.

\section{Basic simulation assumptions}

In actual situations both the information sources and the filters
described in the previous section combine their effects in quite complex
ways. We encounter, in no particular order, information sources of
various type, content and strength, in some cases acting alone, in
others - combined. To elucidate the model effects we shall initially
focus on drastically simplified systems, in which we would show the
effects of the \textbf{repeated application of the same filter to
the same information source distribution} $S_{i}(\theta)$, for a
range of starting belief distributions $X_{j}(\theta,0)$  
(where the index $j$ denotes individual agents).
The aim
of this exercise is to show if particular filters (no filter, confirmation
bias, PMR) lead to stable belief distributions, polarization, emotional
involvement etc.

As noted, for the simulations shown in this paper, 
we shall be using the truth-related
form of the information source, $S_{T}(\theta)$, assumed to take
a rather broad Gaussian form, centred at $\theta_{T}=0.6$ and with
standard deviation equal to 0.4. This choice of the information source
distribution is motivated by two reasons. The first is to check if
the simulated society is capable of reaching consensus when the information
source points to a well defined value. The second reason
was to study the effects of the asymmetry. Obviously, it is much easier
for the agents whose initial opinion distribution favours positive
$\theta$ values to `agree' with an information source favouring a
positive $\theta_{T}$. In contrast, the agents starting with belief distributions preferring negative
$\theta$ values, would have to `learn' and to overcome their initial
disagreement and to significantly change their beliefs. 

\begin{figure}
\includegraphics[width=1\columnwidth]{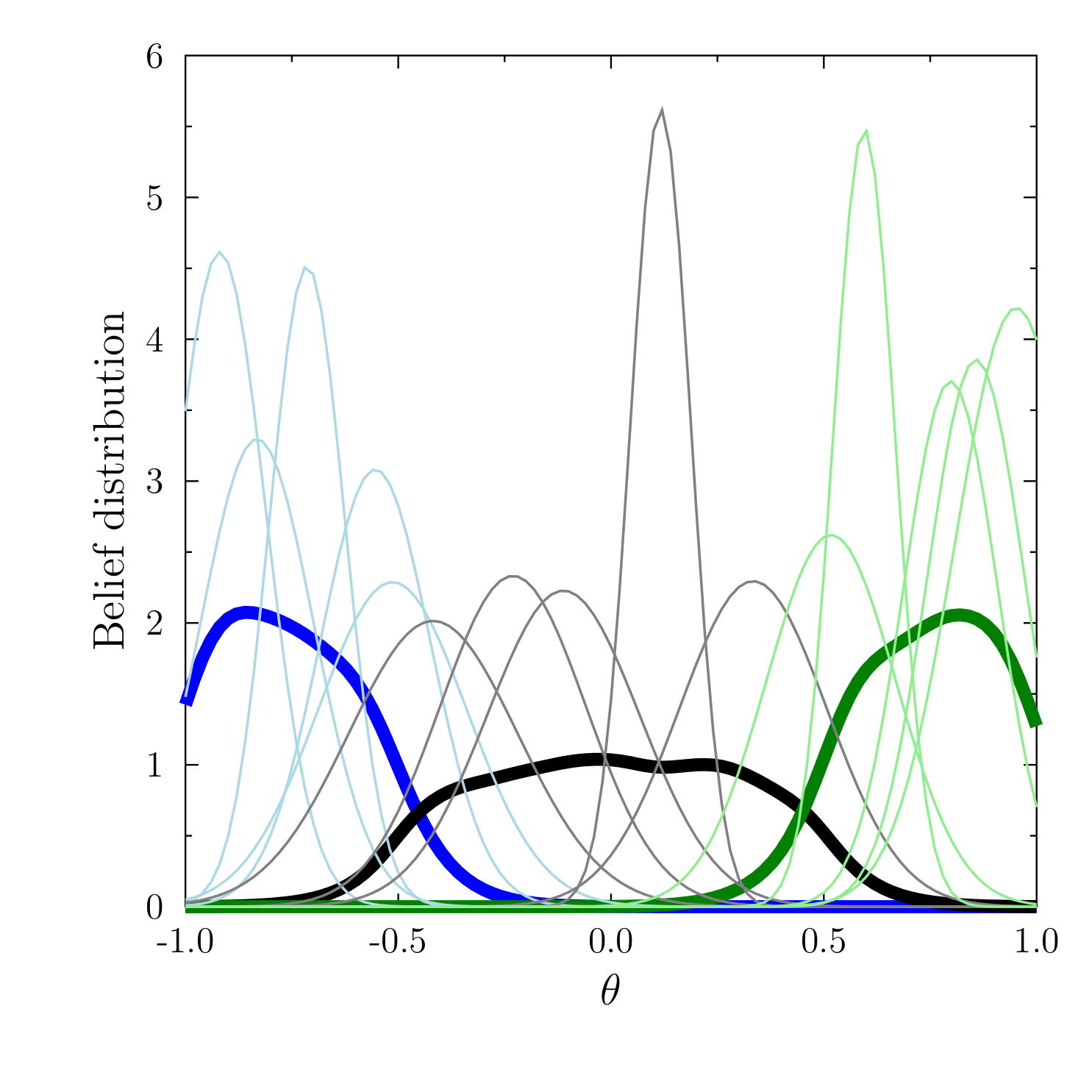}

\caption{Initial belief distributions of three classes of agents. Blue: leftists
(maximum of the initial belief drawn randomly from between -1 and
-0.5). Black: centrists (maximum belief drawn from between -0.1 and
0.5). Green: rightists (maximum belief between 0.5 and 1). Thin, light
lines: examples of the belief distributions of a few individual agents,
differing in their initial centre of the belief $\theta_{0j}$ and
the width of the belief distribution $\sigma_{0j}$. Thick lines:
averaged distributions for each group. \label{fig:Initial-belief-distributions}}
\end{figure}

Each of the agents is initially characterized by
a belief function of a Gaussian form (bounded between $-1$ and +1 and
suitably normalized). The standard deviation parameters for the agents
$\sigma_{0j}$ are drawn from an uniform random distribution limited
between 0.05 and 0.2. 

\textbf{Three separate sets of agents are created and used in the
simulations: leftists, centrists and rightists} (we note here that
these names have no connection with the real world political stances
and \textbf{refer to the position on the abstract $\theta$ axis}).
Each agent community is composed of $N$ agents (in the simulations
we use $N=1000$). 
The leftists
have their initial Gaussian centre values $\langle\theta\rangle_{j}(0)$
drawn from an uniform random distribution bounded between $-1$ and $-0.5$.
The centrist group is formed by agents with $\langle\theta\rangle_{j}(0)$
drawn from between $-0.5$ and $0.5$, and the rightists have $\langle\theta\rangle_{j}(0)$
drawn from between $0.5$ and 1. Figure \ref{fig:Initial-belief-distributions}
shows examples of the agent belief functions for agents from each
of the three groups (thin blue, gray and green lines) and the initial
ensemble averaged belief distributions $X_{G}(\theta,0)$, where $G$
stands for $L$ (leftists), $C$ (centrists) and $R$ (rightists).
These ensemble averages are shown by thick lines. There is some overlap
between the leftist/centrist and centrist/rightist groups, but practically
no overlap between the leftist/rightist groups. 

The simulations discrete time steps. The time is measured in time units in which each agent
in the current group has had a single chance to interact with the
information source or to ignore it (with the respective probabilities
of $p$ and $1-p$). As we shall show in the following sections some
effects become visible after just a few time steps,
but some other become important after thousands or tens of thousands
events. From the point of view of the possible application of model
results to the real life phenomena we should consider the mapping
of the `simulation time' to real hours, days or weeks. In the current
work we focus on the long term behaviour of the system, especially
on the final stable conditions. 

\section{Model results}

\subsection{Case 1: Unfiltered effects of true information}

We shall start the description of the model results with a relatively
simple case, with the aim of showing the effects of some of the simulation
parameters. The first case is based on \textbf{unfiltered} processing of the
`truth-related' information, $S_{T}(\theta)$, as shown in Figure
\ref{fig:sources}. This is, as we have noted, equivalent to the situation
where the information flow is nonspecific (uniform) but the agents
employ a truth-seeking filter of the same form as $S_{T}(\theta)$.
As the $\theta_{T}$ value is positive (equal to 0.6), the most interesting
question is how such information would influence the agents who initially
hold the opposite views (the `leftists'). 

\begin{figure}
\includegraphics[width=1\columnwidth]{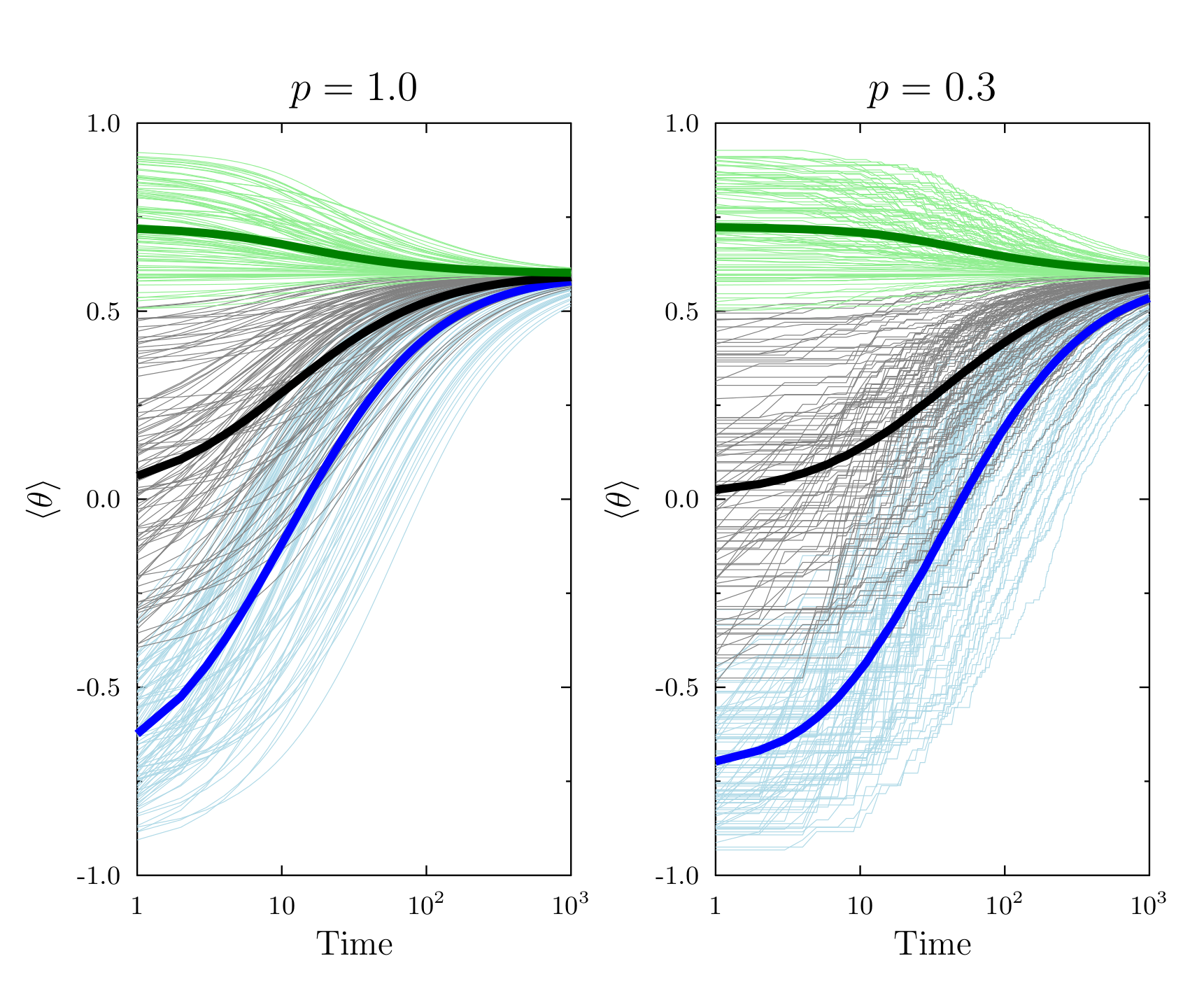}

\caption{\textbf{No filtering applied, perfect memory.} Time evolution of the averages of beliefs of the three groups of agents.
Rightists: green, centrists: black, leftists: blue. Thin line show
evolution of the average belief $\langle\theta\rangle_{j}(t)$ for
individual agents $j$. Thick lines show the evolution of the ensemble
averages for each group of the agents $\langle\theta\rangle_{G}(t)$.
Without filtering, the truth-focused information eventually leads
all the agents to adopt the $\theta_{T}$ centred beliefs. The process
for our choice of $\theta_{T}=0.6$ is, of course, the easiest for
the rightists, who start with beliefs close to this value. However,
eventually all the groups achieve consensus. The case of the leftists
(initially holding views opposing $\theta_{T}$) is quite revealing:
agents with initially large tolerances (large initial $\sigma_{0j}$)
accept the truthful information quickly; agents with very focused
initial opinions (small values of $\sigma_{0j}$) hold on for longer
times and then very quickly join the majority. Simulations that use
the $p=0.3$ value show the periods in which the individual opinions
remain unchanged (indicated by the flat segments of the thin lines).
\label{fig:Time-evolution-of}}
\end{figure}

\begin{figure}
\includegraphics[width=1\columnwidth]{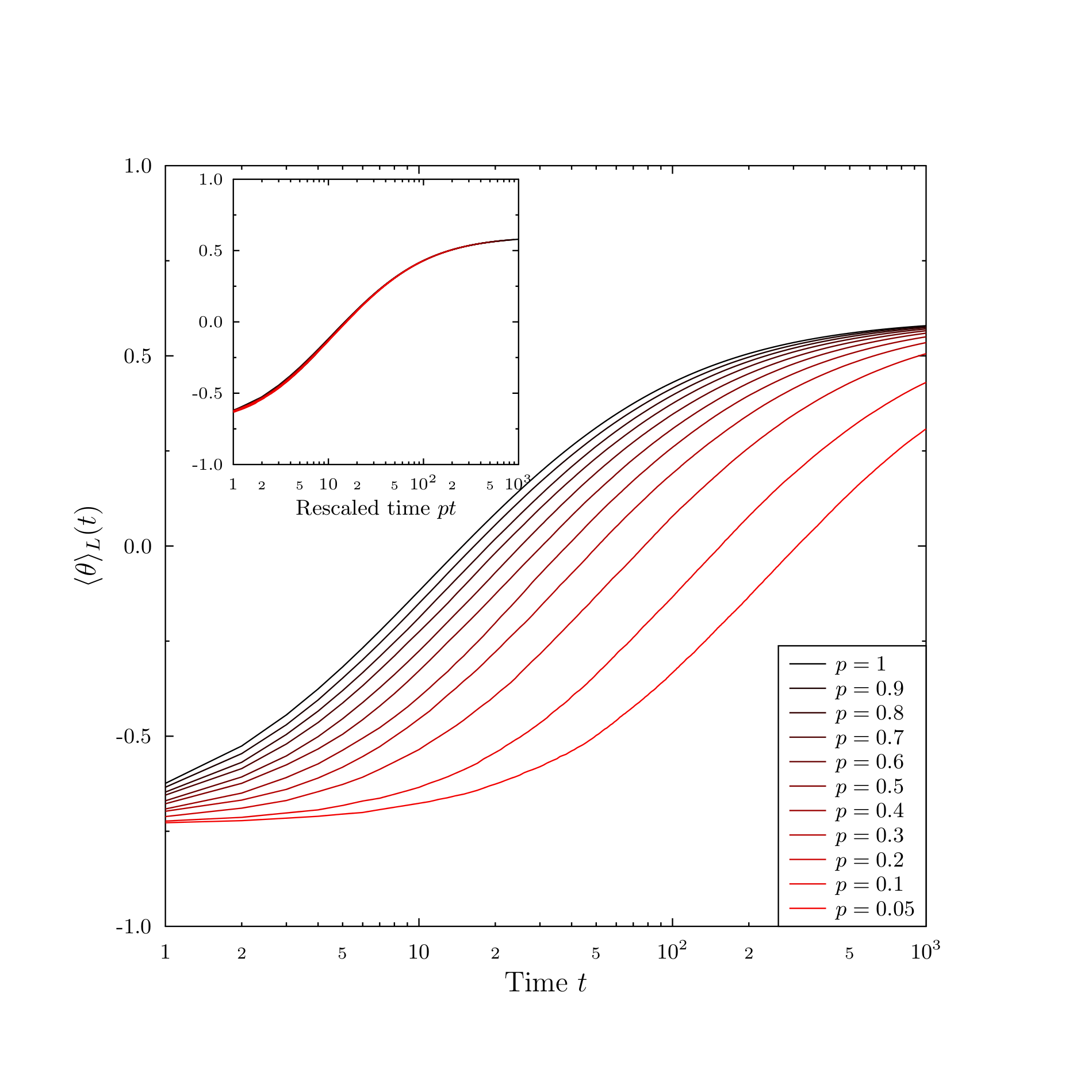}

\caption{\textbf{No filtering applied, perfect memory.} Evolution of the average mean value of the belief $\langle\theta\rangle_L(t)$ 
over the group of `leftists' due to `truth-related' information stream.
Time is measured in interactions per agent. Black-red curves represent
various values of the $p$ parameter value. Decreasing the probability
of information-carrying encounters (smaller $p$) makes the evolution
of the beliefs slower. The pure Bayesian evolution ($p=1$) very quickly
(in less than a 1000 time steps) 
leads to the distribution of beliefs centred around $\theta_{T}$.
For $p<1$ the evolution of $\langle\theta\rangle_L(t)$ is stretched in time proportionally to $p$.
Rescaling time to $t'=pt$ shows the invariant shape of the evolution (inset). \label{fig:slowdown}}
\end{figure}

The speed with which the agents converge at the true value consensus
depends on the significant information processing probability $p$.
Figure \ref{fig:Time-evolution-of} presents the time evolution of
individual agents' average beliefs ($\langle\theta\rangle_{j}$, thin
lines) and the ensemble averages $\langle\theta\rangle_{G}$ for the
three agent groups.
We start with agents characterized by perfect memory ($m=1$).
The time
evolution of the average $\langle\theta\rangle_{j}$ for $p<1$ looks
qualitatively different than in the case of $p=1$.
They exhibit a step-like structure, due to `freezing' of beliefs if no processing takes place. However, 
the ensemble averages are quite
similar for $p<1$ and $p=1$. 
Figure \ref{fig:slowdown} shows the dependence of the time
evolution of the average belief for the whole leftist group $\langle\theta\rangle_{L}$
on the value of the parameter $p$ (note the logarithmic scale of
the time axis). In fact, a simple rescaling of the time axis to $t'=pt$
(shown in the inset) shows that the evolution is really a simple slowdown
due to inactivity periods, when no information is processed. Thus
for the perfect memory (i.e. for $m=1$), the role of $p$ is rather
trivial. It becomes more important when the `idle' times are used
to partially reset the individual uncertainty. 

\begin{figure}
\includegraphics[width=1\columnwidth]{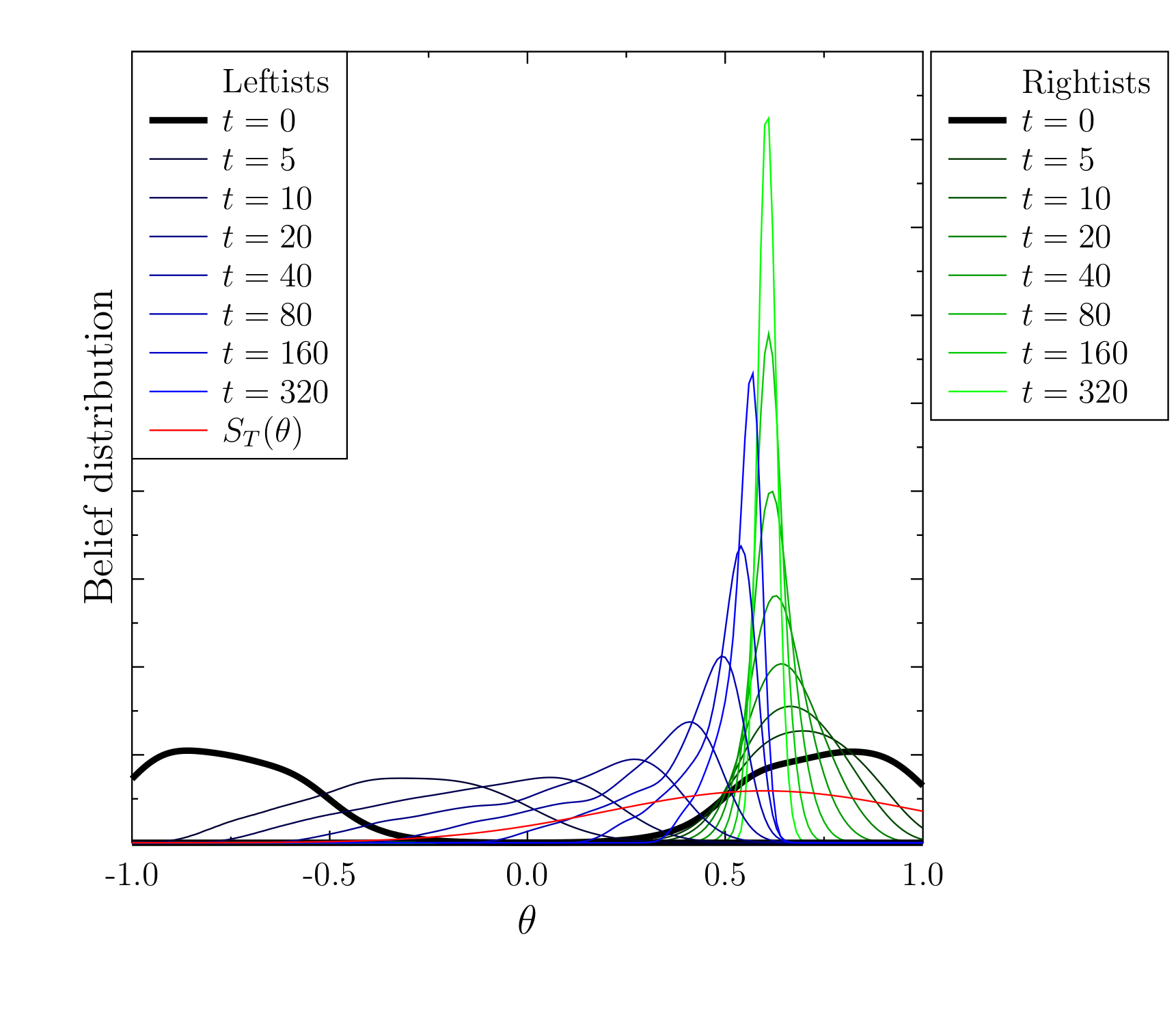}
\caption{\textbf{No filtering applied, perfect memory.} Time evolution of the ensemble averaged belief distribution over the
groups of leftists and rightists due to `truth-related' information
stream. Time is measured in interactions per agent. Black-blue and
black-green curves: averaged belief $\bar{X_{G}}(\theta)$. Red curve
shows, for comparison, the original information distribution $S_{T}(\theta)$.
Simulations use $p=1$ value. As the time passes, the beliefs transform
from the starting polarized distributions (thick black lines) and
converge to the `true' value of $\theta_{T}=0.6$. \label{fig:Evolution1}}
\end{figure}

 The truth-focused information flow eventually
convinces all the agents to believe in the `true' value of $\theta_{T}=0.6$,
regardless of their initial positions. The
process is the fastest for  agents with relatively broadminded
beliefs (high $\sigma_{0j}$). For the agents with initial very narrow
belief distributions the transition is shifted to later times --
but then it is almost instantaneous (typical for Bayesian update of
single valued probabilities rather than distributions). The changes
in the form of the belief distribution consist of a more or less gradual
`flow' of beliefs from the original form to a belief centred around
the maximum of $S_{T}(\theta)$. This is well illustrated by Figure
\ref{fig:Evolution1}, which presents the time evolution of the ensemble
average belief for the leftist and the rightist groups. 

\begin{figure*}
\includegraphics[width=1\textwidth]{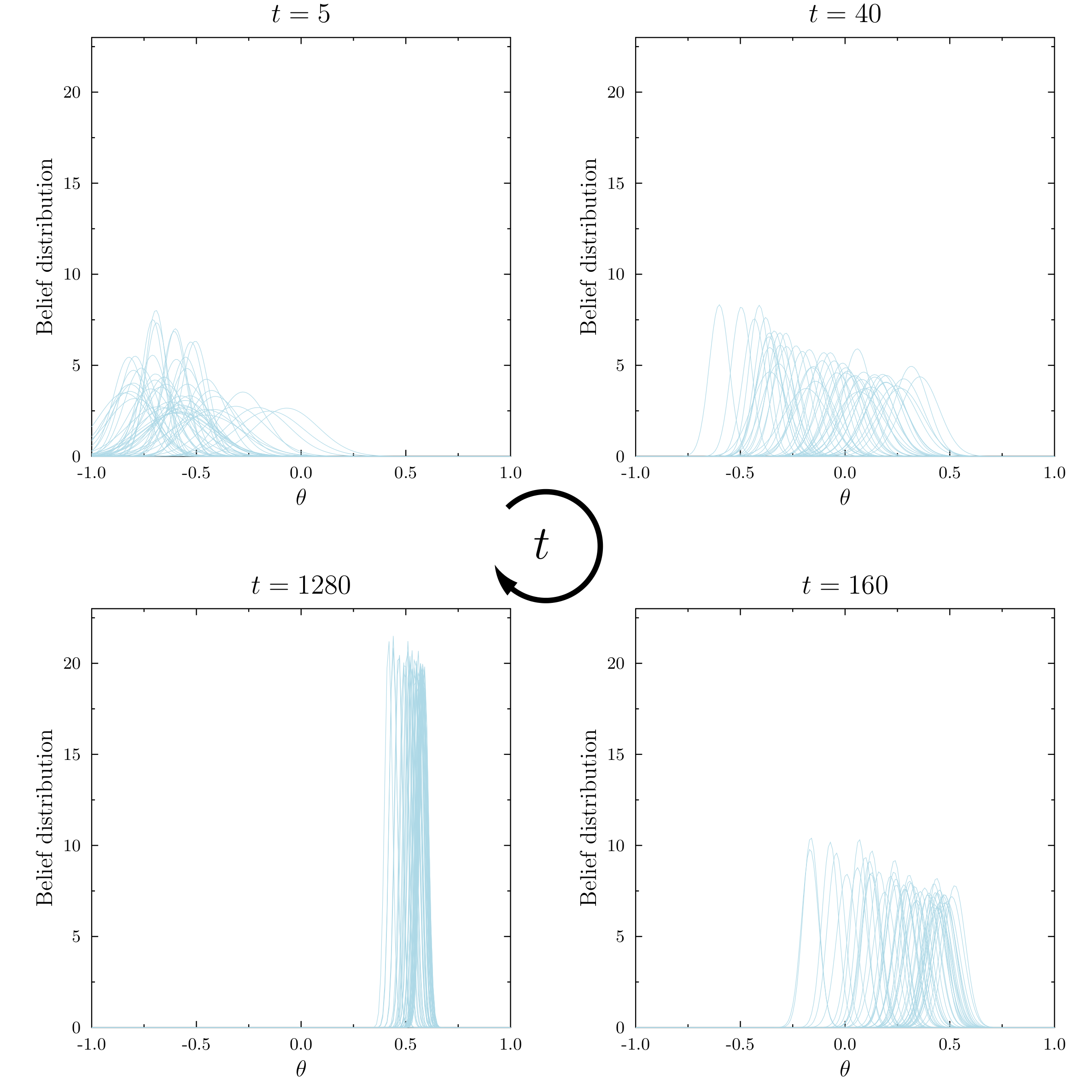}

\caption{\textbf{No filtering applied, perfect memory.} Snapshots of examples of the individual agent belief distribution
functions for the leftist agents under the influence of the unfiltered
$S_{T}(\theta)$ information source. For the simulations $p=0.3$
and $m=1$ (that is, perfect memory is assumed and no indeterminacy
of beliefs occurs). As expected, the individual beliefs move towards
the true value of $\theta_{T}=0.6$, at the same time becoming increasingly
narrow. There is only a partial overlap of the individual opinions.
The snapshots are taken at $t=5,40,160$ and $1280$ (shown in clockwise
order). \label{fig:SnapshotsC1m1}}
\end{figure*}

\begin{figure*}
\includegraphics[width=1\textwidth]{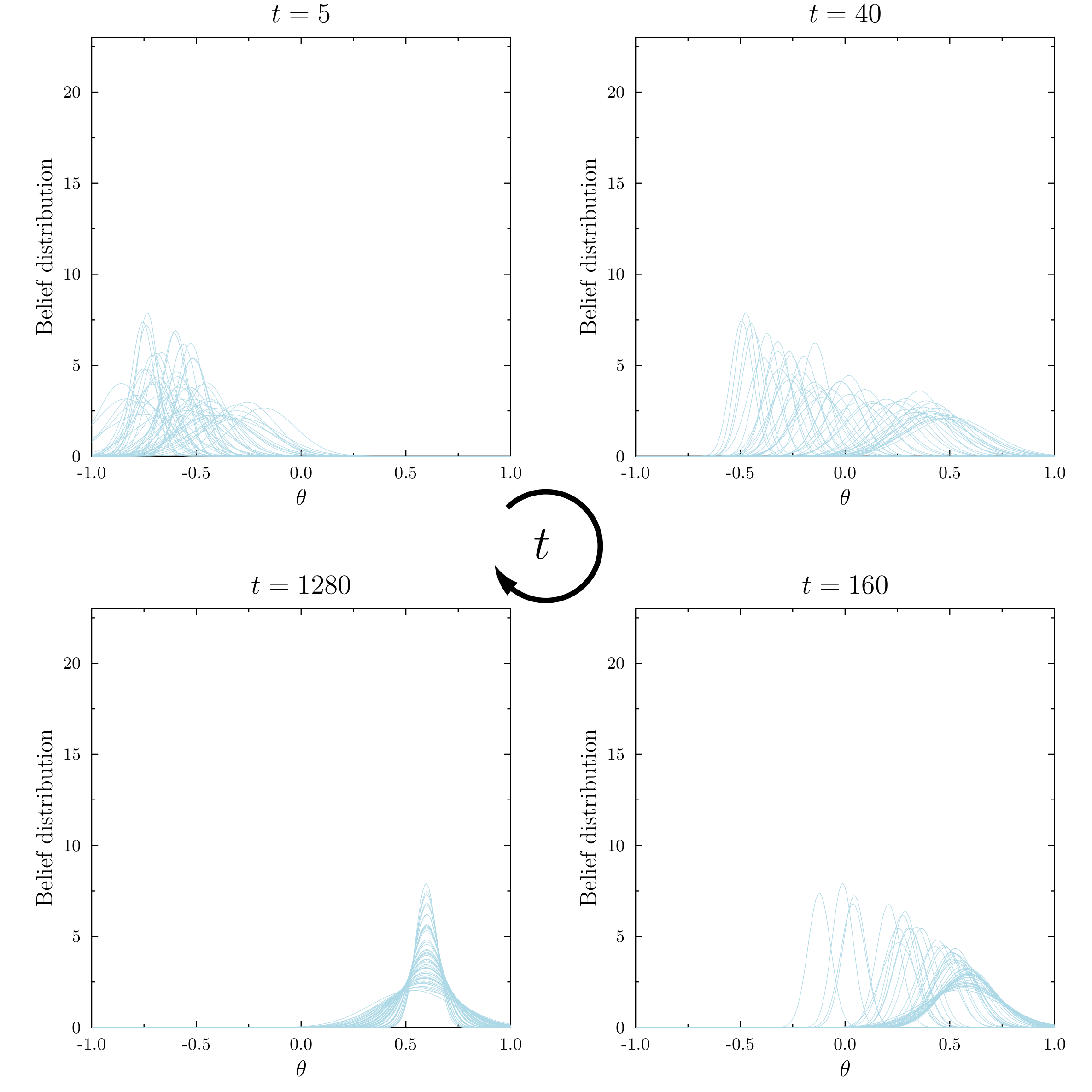}

\caption{\textbf{No filtering applied, reset of belief indeterminacy.} Snapshots of examples of the individual agent belief distribution
functions for the leftist agents under the influence of the unfiltered
$S_{T}(\theta)$ information source and with the indeterminacy reset
present. For the simulations $p=0.3$ and $m=0.5$. In this case,
not only do the individual beliefs move towards the true value of
$\theta_{T}=0.6$, but they also become almost fully overlapping.
The snapshots are taken at $t=5,40,160$ and $1280$ (shown in clockwise
order). \label{fig:SnapshotsC1m05}}
\end{figure*}

To understand the effects of the memory parameter $m$, it is illustrative
to study the effects of the indeterminacy reset on the evolution of
the \textbf{individual }opinion distributions $X_{j}(\theta,t)$.
As shown in Figures \ref{fig:SnapshotsC1m1} and \ref{fig:SnapshotsC1m05}
the relaxation of the indeterminacy introduced by imperfect memory
factor $m<1$ leads to a qualitatively different final form of the
individual belief distributions. Instead of a set of narrow, delta-like
functions grouped close to the $\theta_{T}$ value typical for $m=1$,
the existence of belief relaxation leads to distributions of width
comparable to the original values of $\sigma_{0j}$ centred exactly
at $\theta_{T}$. Thus, while the final ensemble average may be similar,
the underlying structure of the individual beliefs is quite different.

\begin{figure}
\includegraphics[width=1\columnwidth]{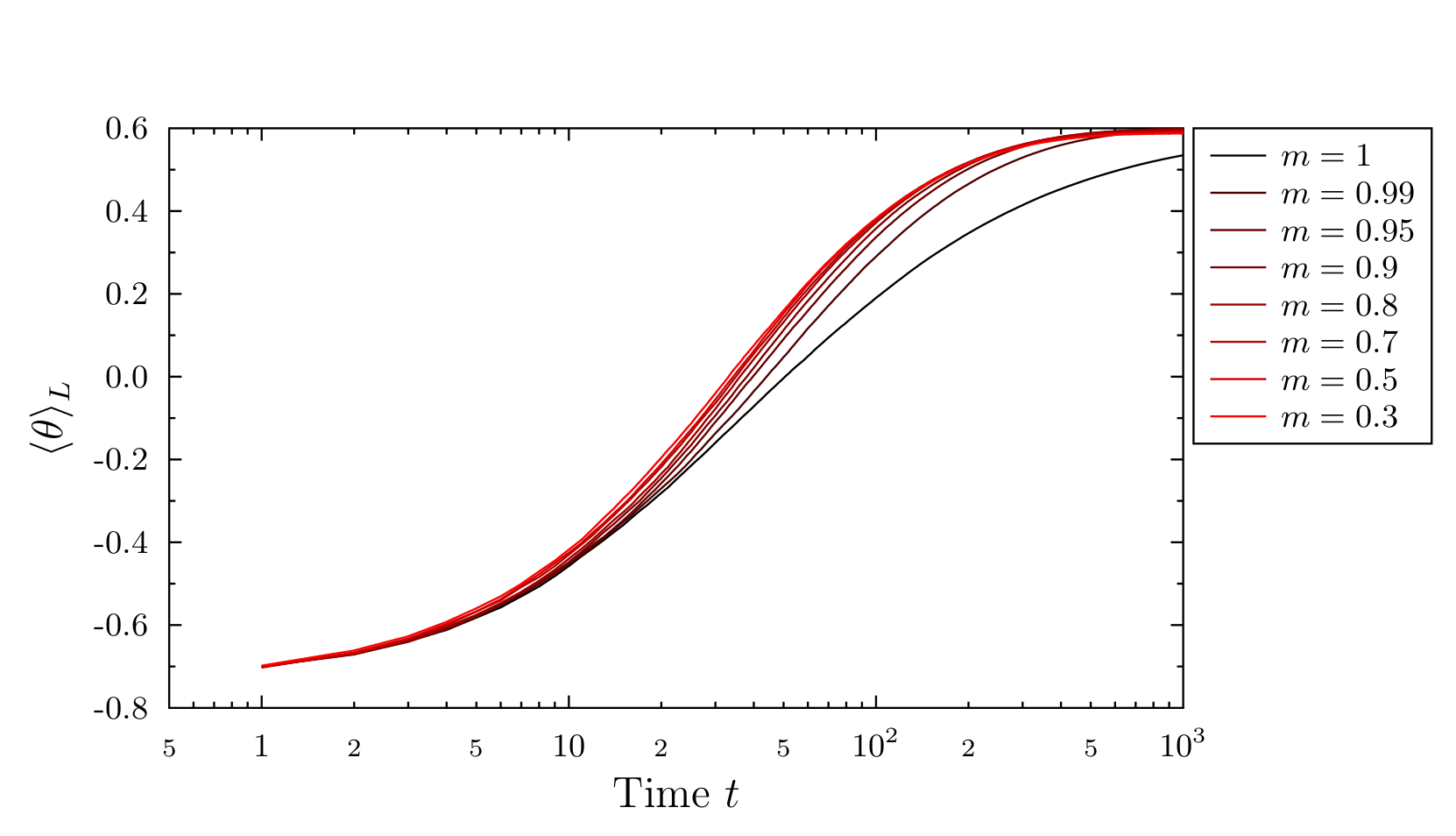}

\caption{\textbf{No filtering applied, overview of memory effects.} Time evolution of the leftist group ensemble average $\langle\theta\rangle_{L}(t)$
for $p=0.4$ and various values of the memory factor $m$. Even a
relatively small memory loss ($m=0.99$) speeds up the transition
of $\langle\theta\rangle_{L}$ to the true value $\theta_{T}$. In
other words, agents with belief distributions which are (however seldom)
reset to more `broadminded' values are more likely to learn from the
`truth-related' information source. \label{C1PxaverNOscaling}}
\end{figure}

In the case of lack of filtering, the effects of the indeterminacy
reset on group averages are rather subtle. 
For a given value of $p$, decreasing the memory factor leads
to a small, but observable shortening of the time scale of reaching the truth-based consensus (Figure~\ref{C1PxaverNOscaling}). 
It is interesting that even a very small admixture of uncertainty reset
($m=0.99$ instead of $m=1$) significantly influences the evolution of the
group averaged belief $\langle\theta\rangle_L$.  

\subsection{Case 2: Individual confirmation bias filter of true information,
perfect memory ($m=1$).}

In the previous section we have shown that under the influence of the truth-related information, without filtering,
all the agents eventually converge their beliefs on the true value
suggested by the information source. This is not surprising, as the
process is a simple, repetitive, application of a Bayesian
belief modification. We turn now to the issue of the effects of filtering
of the information sources.

We shall start with the individual agent based \textbf{confirmation bias filter}.
There are two reasons for this choice. The first is that confirmation
bias is widely recognized in psychological literature, so it `deserves'
a thorough treatment in the ABM framework. The second reason is a
relative simplicity of the filter effects. Suppose that the information
flow on which the filter acts is non-specific (i.e. uniform). If the
initial belief distribution is given by a Gaussian function with the
standard deviation $\sigma$, then the application of the same function
acting as the likelihood filter would lead to the posterior belief
in the Gaussian form, but with $\sigma$ decreased by a factor of
$\sqrt{2}$. A repeated information processing would eventually lead
to a Dirac delta-like belief distribution. In other words, repeated
application of confirmation bias narrows and freezes one's own opinions.
Once should, therefore, expect that the confirmation bias filter should
diminish the effects of specific information sources, such as the
truth-related source $S_{T}(\theta)$.

The simulation setup for the case of confirmation bias filtering of
true information is relatively simple. At every time step, with probability
$p$ each agent uses its current belief $X_{j}(t)$ as the pure filter.
In this case the final likelihood function is defined by 
\begin{equation}
FL(\theta)=(fX_{j}(\theta,t)+(1-f)U)S_{T}(\theta),
\end{equation}

where we use the filter effectiveness $f$ as parameter. As before,
with probability $1-p$, the agent does not process the information.
In this section we focus on situations $m=1$, when the agent simply
retains its previous belief. In what follows, we use a fixed value
$p=0.3$. The crucial parameter in Case 2 is the filter effectiveness
$f$.

\begin{figure*}
\includegraphics[width=1\textwidth]{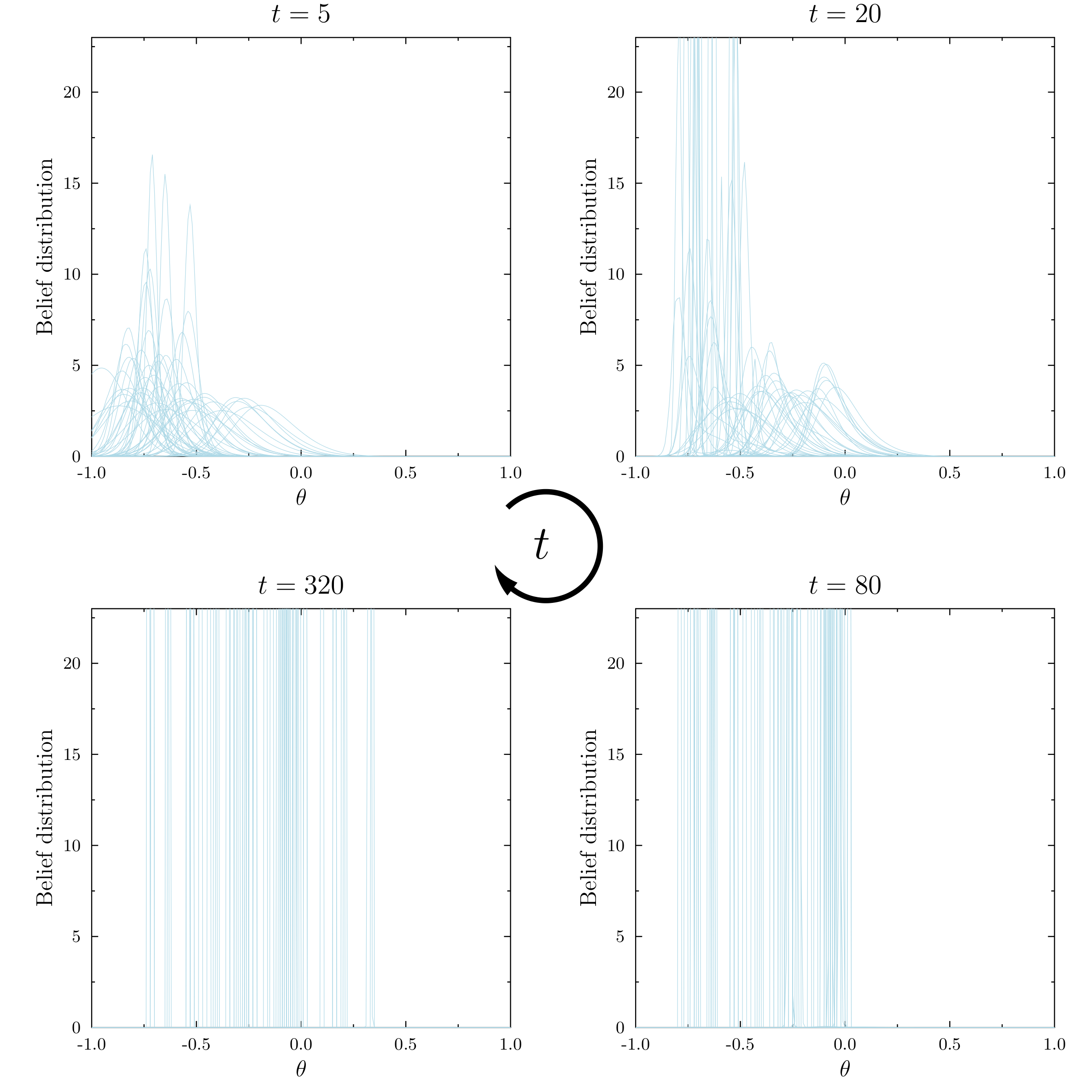}

\caption{\textbf{Confirmation bias filtering, perfect memory.} Time evolution of individual belief distributions of selected leftist
agents using confirmation bias filtering of truth-related information
for $f=0.2$, $p=0.3$ . The distributions are for a subset of randomly
selected leftist agents. As the time progresses (clockwise) the individual
belief distributions shift to higher $\theta$ values and become increasingly
narrow. The first process is the influence of the new information
favouring $\theta_{T}=0.6$, the second is the result of the confirmation
bias. Quite quickly (much faster than in the case of unfiltered information
processing), all agents evolve to delta-like belief distributions
(the $t=80$ panel). For $f$ values greater than 0.05 the processes
of narrowing of the individual beliefs dominates over their shift
towards the true value for a large number of the agents. \label{fig:4panels}}
\end{figure*}

Figure \ref{fig:4panels} presents four snapshots of the evolution
of the individual belief distributions $X_{j}(\theta)$ for the leftist
group. The individual distributions change under the competing influences
of the confirmation bias filter (progressively narrowing the belief
distributions) and the information (shifting the beliefs towards higher
values of $\theta$). 

\begin{figure*}
\includegraphics[width=0.9\textwidth]{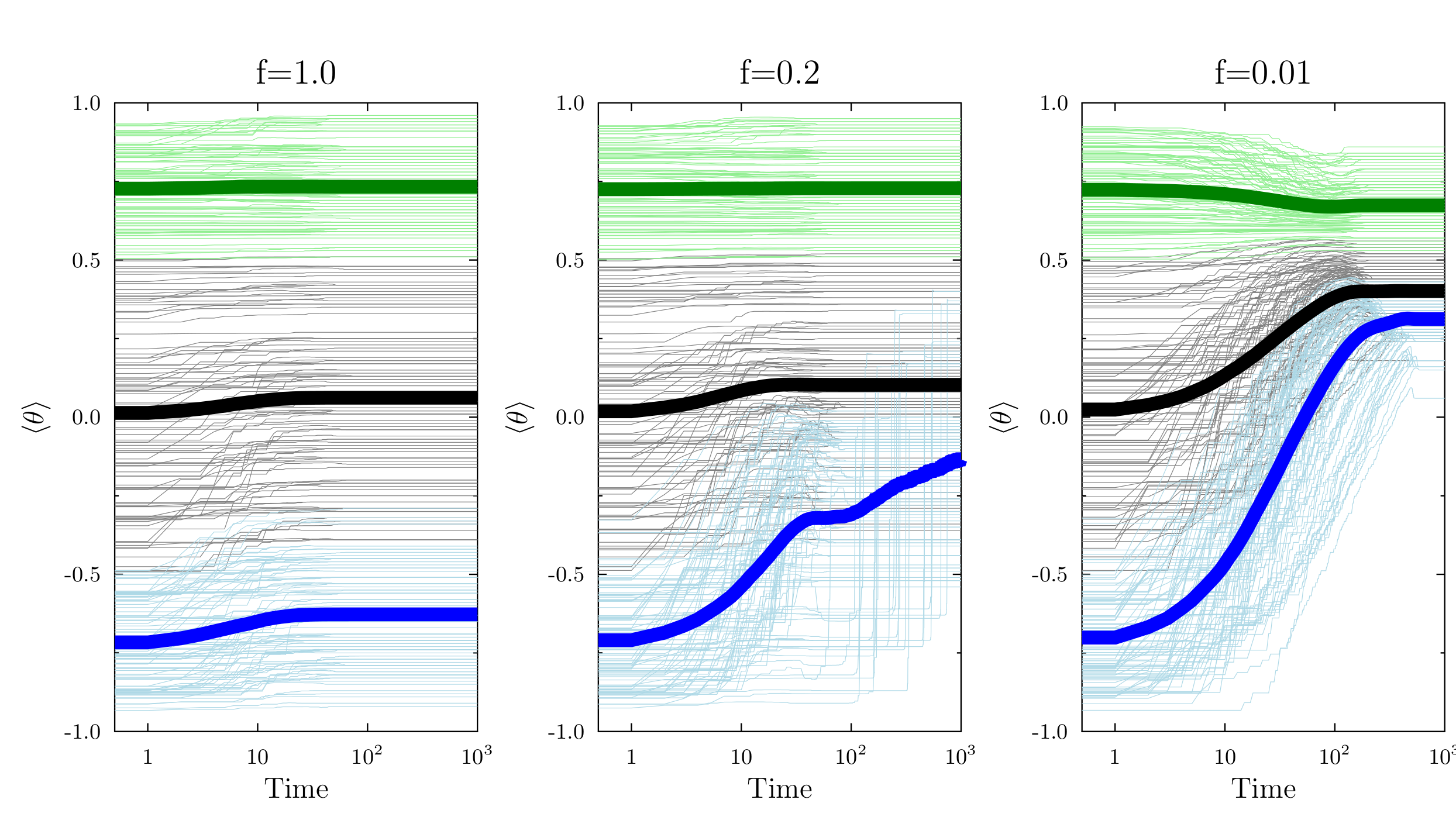}

\caption{\textbf{Confirmation bias filtering, perfect memory.} Time evolution of average beliefs $\langle\theta\rangle_{j}$ (thin
lines) and the averages $\langle\theta\rangle$ (thick lines) for
the three groups of agents using confirmation bias, for three values
of the filter effectiveness $f=1,0.2$ and $0.01$. The value of the
information processing probability is $p=0.3$. Decreasing the effectiveness
of the confirmation bias filter delays the time at which the individual
opinion distributions become fixed and delta-like, shown in the figure
as think horizontal lines. In some cases we observe jumps in the opinion,
typical for discrete Bayesian updates.\label{fig:3evolutions}}
\end{figure*}

The relative importance of these two factors
depends on the value of the filtering effectiveness factor $f$. If
a pure form of the filter is used ($f=1$) the individual beliefs
coalesce to delta form in less than 10 time steps and the average
beliefs of the three groups remain practically unchanged (left panel
in Figure \ref{fig:3evolutions}). Thus, despite the availability
of true information, the centrist and leftist groups keep their beliefs.
For much smaller, but still non-negligible value of $f=0.2$ (corresponding to Figure~\ref{fig:4panels}), we observe
some change (more pronounced for the leftist group, where the dissonance
between the initial views and the true information is the largest,
middle panel in Figure \ref{fig:3evolutions}). 

\begin{figure*}
\includegraphics[width=0.9\textwidth]{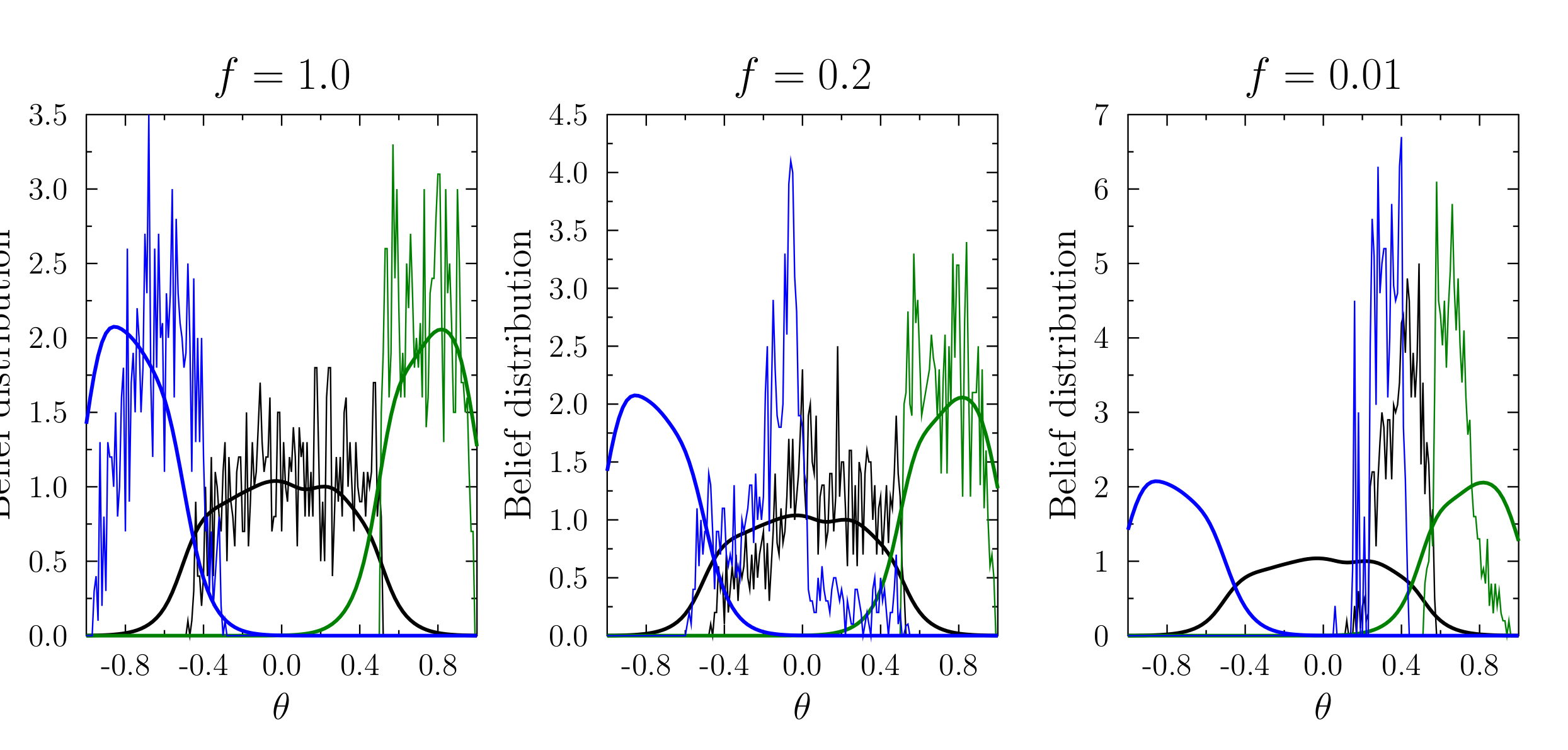}

\caption{\textbf{Confirmation bias filtering, perfect memory.} Averaged distributions of agent beliefs in the three groups. Thick,
smooth lines: initial distributions, thin lines: distributions after
10000 steps. Results for three $f$ values are shown $f=1$, $f=0.2$
and $f=0.01$. \label{fig:Averaged-distributions-of}}
\end{figure*}

It is only for very small values of $f$ that the final distributions
of beliefs begin to converge towards the truth-related consensus.
Even for $f=0.02$ there is a sizeable gap between the rightists and
the centrists and leftists. Figure \ref{fig:Averaged-distributions-of}
presents the shape of the ensemble averaged belief distributions
for each of the three groups at a very late time $t=10000$. 

\begin{figure}
\includegraphics[width=1\columnwidth]{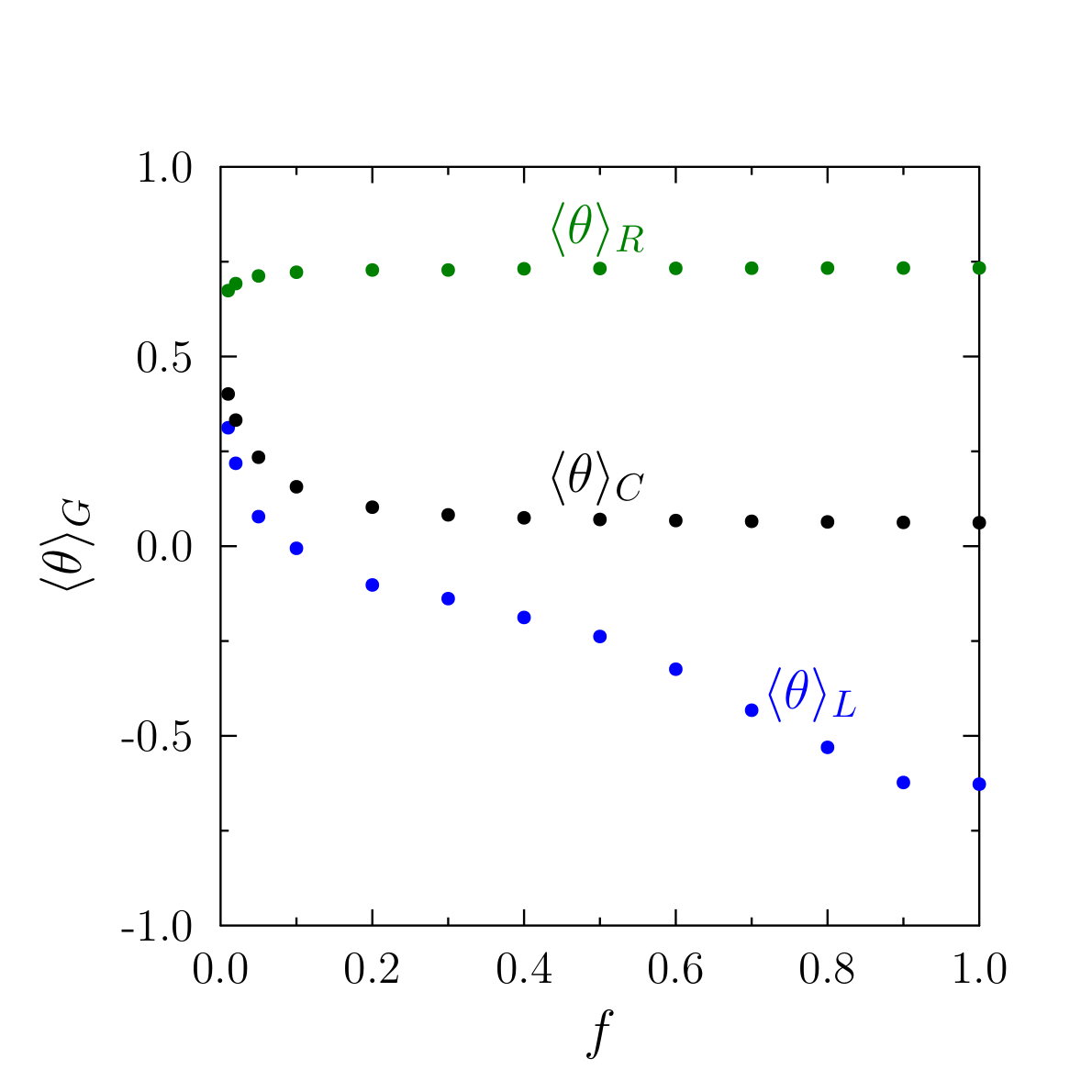}

\caption{\textbf{Confirmation bias filtering, perfect memory.} Dependence of the final value of $\langle\theta\rangle_{G}$ for the
three groups as functions of filtering effectiveness $f$ for the
confirmation bias filter. Note that the convergence of opinions near
the true value requires very weak filtering $(f<0.5$). \label{fig:Dependence-of-the}}
\end{figure}

Figure \ref{fig:Dependence-of-the} presents the dependence of the
ensemble averaged values of the average belief for each of the three
groups on the filtering effectiveness $f$. For $f$ close to 1, the
truth-related information is almost totally filtered out by the confirmation
bias, the agents quickly evolve to fixed, delta-like belief distributions.
For medium values ($0.3<f<0.9$) the rightists and the centrists show
no effects, but the leftists are gradually `convinced' to shift their
opinions somewhat towards positive value. For small values of the filtering effectiveness
($f<0.1$) the opinions of the three groups begin to converge, but
getting close to consensus requires very small values of $f$ (on
the order of 0.02 or less).

\subsection{Case 2a: Individual confirmation bias filter of true information,
reset of beliefs due to imperfect memory ($m<1$). }

\begin{figure*}
\includegraphics[width=1\textwidth]{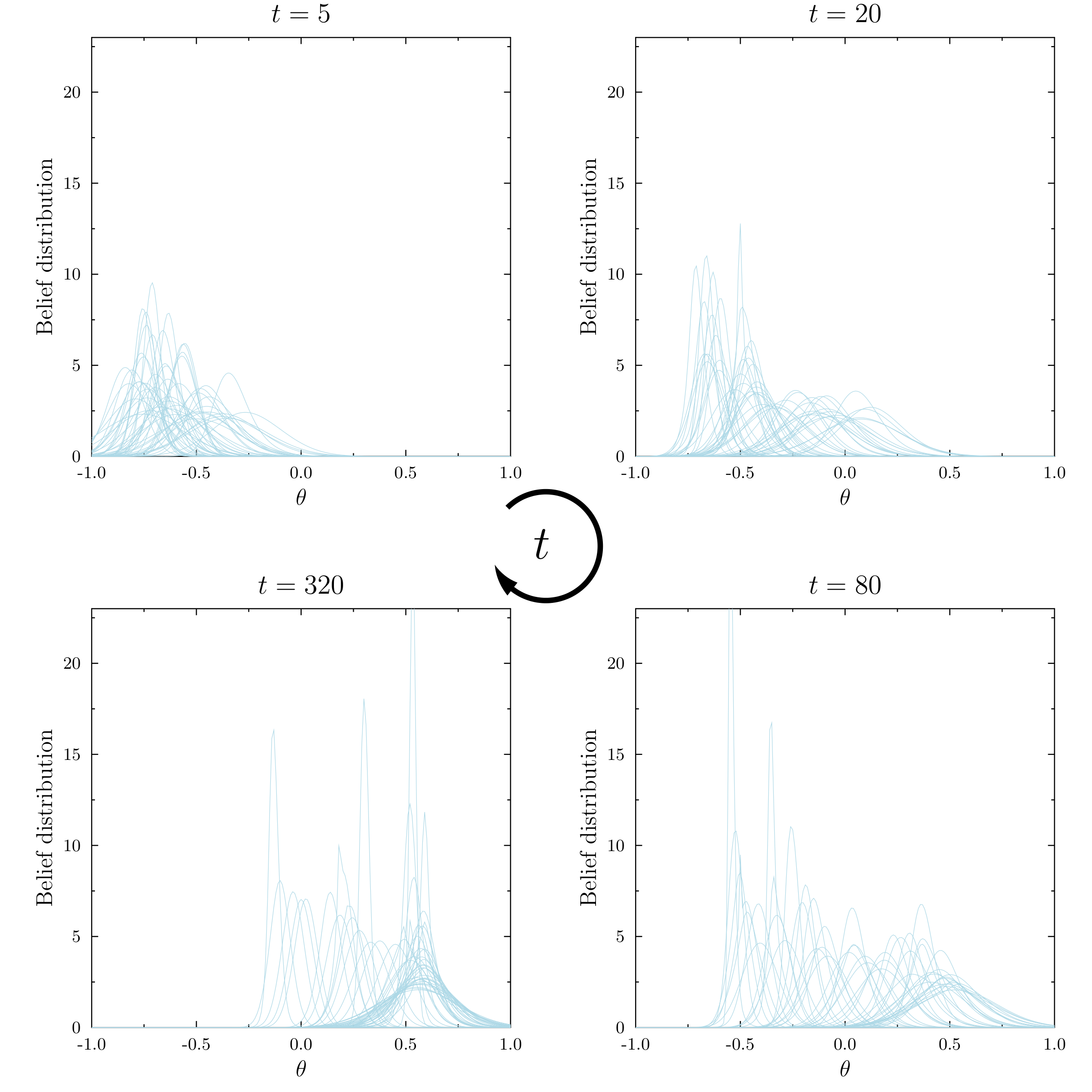}

\caption{\textbf{Confirmation bias filtering, reset of belief indeterminacy.} Snapshots in the evolution of individual belief distributions of selected
leftist agents using confirmation bias filtering of truth-related
information for $f=0.2$, $p=0.3$ with the memory imperfection factor
$m=0.5$. The distributions are for a subset of randomly selected
leftist agents. As the time progresses (clockwise) the individual belief
distributions shift to higher $\theta$ values but remain rather broad-shaped
(as are the original distributions). Much greater number of agents move
to the true belief $\theta_{T}$. Eventually all agents would reach
consensus at this value. \label{fig:4panels2a}}
\end{figure*}

The confirmation bias filter very quickly
leads to extreme narrowing of the individual belief distributions
(for the fully effective case $f=1$ this happens after a few tens
of interactions). This suggests that the inclusion of the broadening mechanisms
might have more significant effect than in the case of unfiltered
information processing. Indeed, setting $m=0.5$ changes the evolution
of the individual beliefs dramatically, as we can see from Figure
\ref{fig:4panels2a} (which corresponds to the `perfect memory', $m=1$
case in Figure \ref{fig:4panels}). In situation when the beliefs
are affected by the indeterminacy reset (which, we remind, does
not change the current individual average belief), they are much more modifiable by
the information source.

\begin{figure}
\includegraphics[width=1\columnwidth]{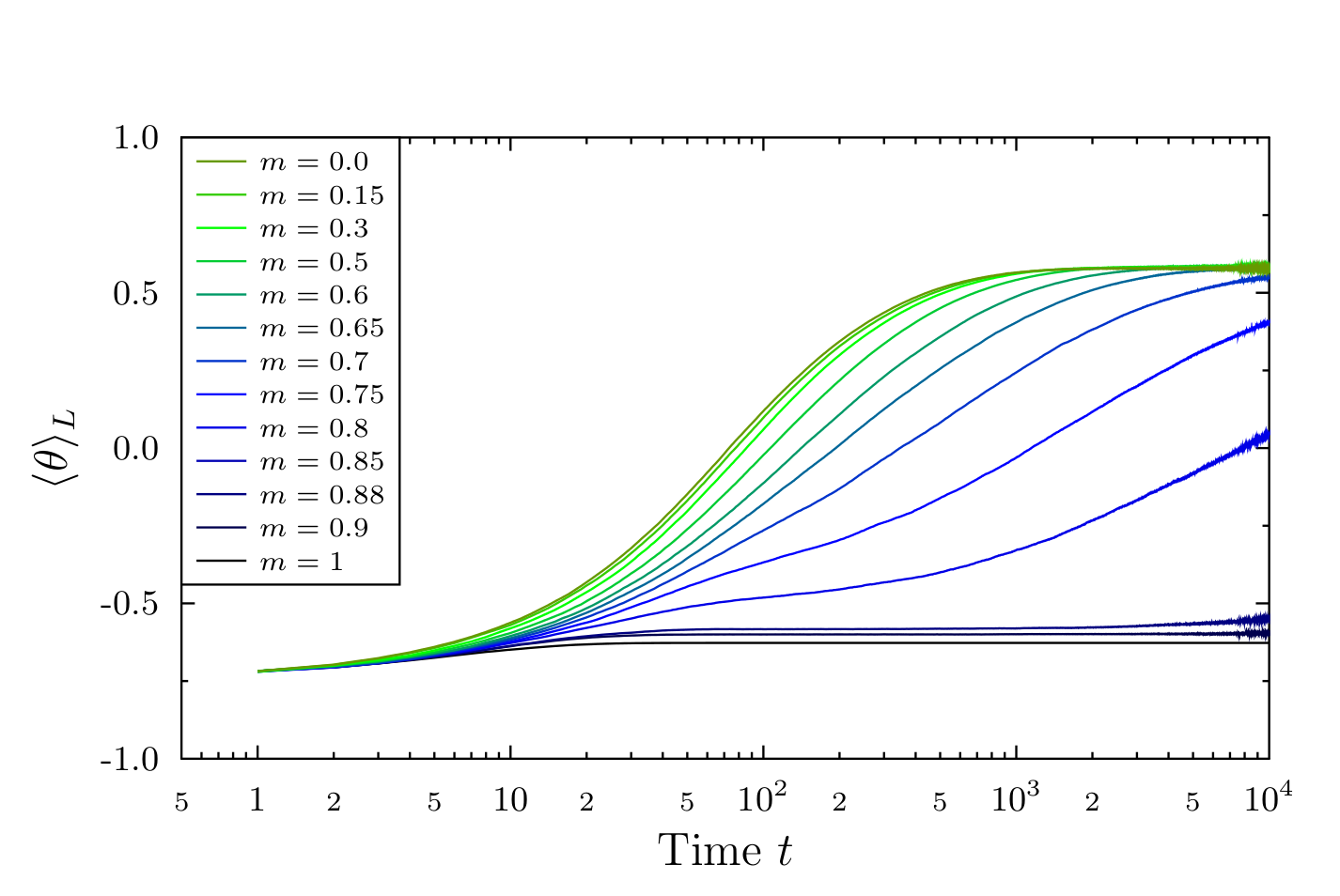}

\caption{\textbf{Confirmation bias filtering, memory effects.} Time dependence of the value of $\langle\theta\rangle_{L}$ for the
leftist group for various values of the memory parameter $m$, for
$f$ equal to 1. Reducing the value of $m$ changes the evolution
of the individual beliefs, and, in consequence, the group average
$\langle\theta\rangle_{L}(t)$: for $m$ smaller than certain value
(significant broadening), all agents become `convinced' by the information
source and accept the $\theta_{T}$ as the centre of their belief
distributions. The conviction process is the fastest for the lowest
values of $m$. On the other hand, for $m>0.9$ the agents' belief
distributions remain frozen, which means that the whole system would
exhibit significant polarization. \label{fig:m-timedependence}}
\end{figure}

In the case of the $S_{T}(\theta)$ information source, the effects
of the memory imperfection are most clearly seen by the behaviour of
the leftist group,
because this group is the furthest from the `true' value $\theta_T$. 
The change introduced by the indeterminacy reset
is best visualized when we look at the time evolution of the
group ensemble average beliefs $\langle\theta\rangle_{L}$ (Figure~\ref{fig:m-timedependence}). The presence
of the indeterminacy reset due to imperfect memory causes the individual
opinion distributions to retain some component of broader beliefs
and facilitates their shift due to the information source influence.
As the process of  reaching the consensus looks similar for all groups, this would lead to a
global consensus centred at the $\theta_{T}$ value. The process $\langle\theta\rangle_L(t) \rightarrow \theta_T$
is quite fast, on the order of a few hundreds of time steps when $m\leq0.3$,
but slows down for higher values of $m$. Above $m\approx0.9$ (i.e.
for an almost perfect memory) the narrowing of the individual opinion
distributions dominates and the group average remain close to their
initial values. 
In plain words, when the agents are allowed to become extremely close-minded due to the confirmation bias, the truth-related information has
limited effect, and the initial assumed polarization between the leftists,
centrists and rightists remains unchanged. 

\begin{figure}
\includegraphics[width=1\columnwidth]{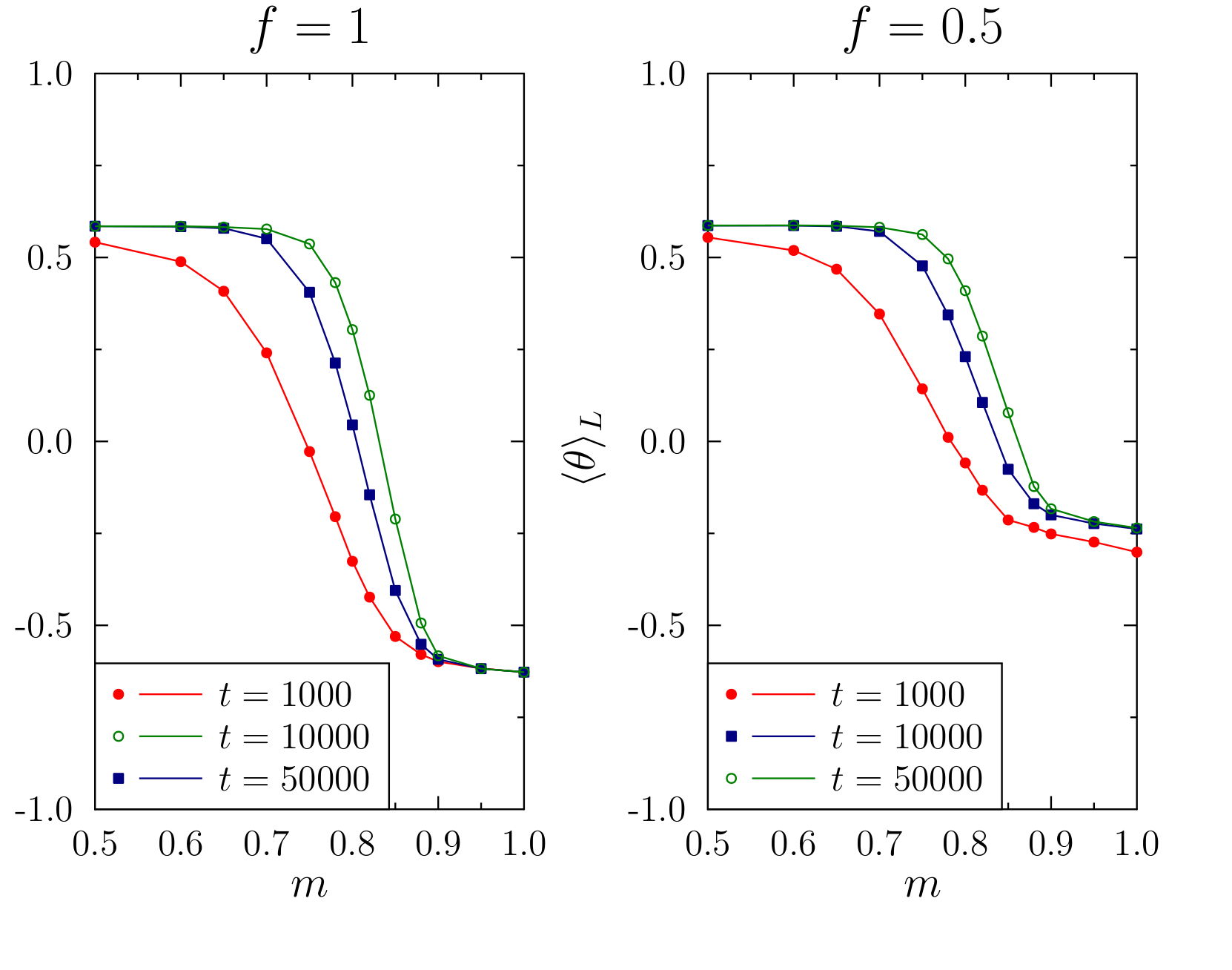}

\caption{\textbf{Confirmation bias filtering, memory effects.} Dependence of the value of leftist group ensemble average $\langle\theta\rangle_{L}(t)$ as function of the memory parameter $m$, for two values
of the filtering effectiveness $f=1$ and $f=0.5$, and for three time
values, $t=1000$, 10000 and 50000 steps. For small $m$ values,
the group average converges on the true value $\theta_{t}=0.6$. For
large $m$ values (better memory, i.e. lesser role of indeterminacy
reset) the beliefs remain on the left side of the opinion spectrum.
Increasing the time $t$ at which we measure the $\langle\theta\rangle_{L}(t)$,
makes the transition between two regimes (preserving the original
opinions and accepting the true value) less gradual as function of
the memory fidelity parameter $m$.\label{fig:m-dependence}}
\end{figure}

The transition between the polarized state at large enough $m$ and
the consensus, for smaller $m$ values, is rather abrupt. Figure \ref{fig:m-dependence}
presents the dependence of the $\langle\theta\rangle_{L}(t)$ values
on $m$, for two values of the filter effectiveness $f=1$ and 0.5
for three time snapshots, $t=1000$, 10000 and 50000. Increasing the
time leads to a step-like transition between conditions preserving
the polarization and those leading to the consensus. 

We recall here
the brief discussion of the topic of mapping the simulation time to
the real world units. Obviously, if we consider as events the cases
when a person encounters really new, significant information (e.g.
listens to a candidate speech at a rally, or a debate, or reads an
important article in the press), then 50000 events is obviously not
realistic. Even a few hundred events (necessary to reach consensus
for very imperfect memory, $m\approx0$) may be questioned. On the
other hand, if we treat the time `between the events' - essentially
the very time in which the memory imperfection and uncertainty reset
would be expected to occur as single entities or, perhaps, a multitude
of them. A partial answer could be provided by psychological research
devoted to the issue of the existence of the indeterminacy of opinion
resets and the associated conditions. 

\subsection{Case 3: Politically Motivated Reasoning filter\label{subsec:Case-3} }

In contrast with the confirmation bias, the PMR filter is assumed
to depend on the current beliefs of the in-group, treated as a whole.
In the simplest version, we assume that any agent knows perfectly
the ensemble averaged belief distribution of its in-group $X_{G}(\theta,t)$,
and uses it as a filter for information processing. The filter is
dynamical, because as the individual agents change their beliefs,
so does the average for the group.  As in the previous
sections we focus on the truth-related information source $S_{T}(\theta)$
and assume that $p=0.3$. Our focus is, therefore, the role of the
filter effectiveness $f$ in the evolution  of the group
belief distributions. 
The current section considers the case of agents with perfect memory ($m=1$). 

\begin{figure}
\includegraphics[width=1\columnwidth]{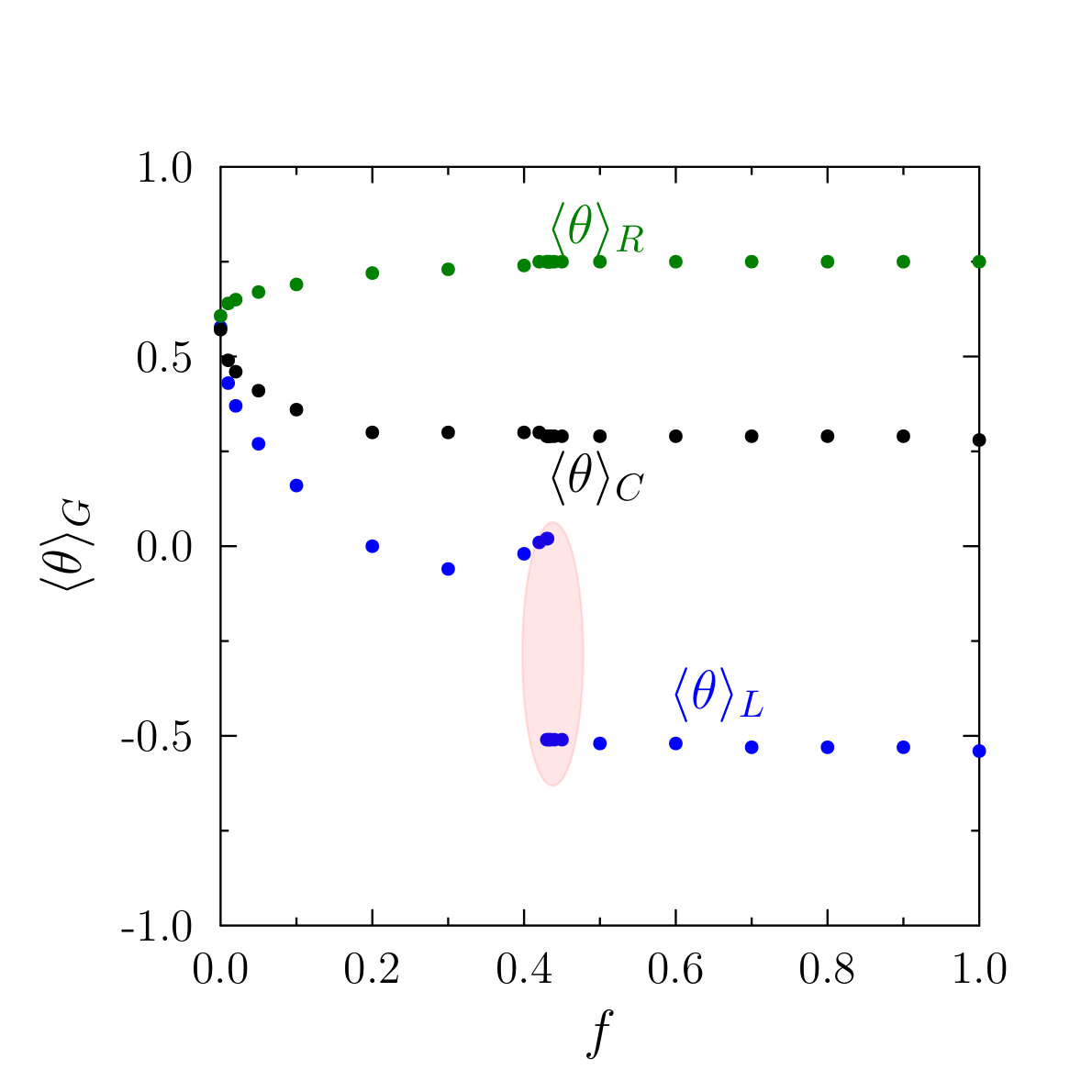}

\caption{\textbf{Politically Motivated Reasoning filtering, perfect memory.} Dependence of the final value of $\langle\theta\rangle_{G}$ for the
three groups, as functions of filtering effectiveness $f$ for the
PMR filter.For $f\gtrsim0.43$ the averages are almost independent
of $f$. At $f\approx0.43$ (marked by the red ellipse), the $\langle\theta\rangle_{L}$
shows a large jump towards the $\theta_{T}$ value, the result effectively
turns the leftists into centrists. For very small values of the filtering
effectiveness ($f<0.1$) opinions of all three groups converge on
the true value $\theta_{t}=0.6$.\label{fig:case3jump}}
\end{figure}

We shall start with Figure \ref{fig:case3jump}, which corresponds
directly to the results for the confirmation bias filter (Figure \ref{fig:Dependence-of-the}).
For very small values of $f$ the averaged beliefs converge on the
true value, as the information source `gets through', 
thanks to the uniform part
of the filter. On the other hand, for $f\approx1$, the PMR filtering
mechanism effectively freezes the group opinions. 
For the two groups
which are initially closer to the true opinion $\theta_{T}$, namely
the rightists $\langle\theta\rangle_{R}$ and centrists $\langle\theta\rangle_{C}$,
the fixed value remains unchanged as we lower $f$, and for very small
values of $f$ it changes gradually,
resembling the behaviour for the confirmation bias filter. 
For the leftists, however, instead
of a continuous change observed in the confirmation bias case we observe a discontinuous transition
at certain value $f_{crit}=0.43$ (for the current set agents and $p=0.3$). 

To understand this discontinuity we have to look into the details of
the evolution of the individual belief distributions. The two following
Figures (\ref{fig:Timef430} and \ref{fig:Timef420}) show examples
of time snapshots of the individual belief distributions $X_{j}(\theta,t)$,
collected for $f$ just above the transition value ($f=0.43$) and below it ($f=0.42$).
The starting point is the same in the two cases. The initial evolution ($t<10$) is driven by the interplay of
the asymmetry of the information source (favouring positive values
of $\theta$) and the PMR filter. It 
leads to formation of two attractors, around which the individual
agents group: one close to the upper end of the original leftist domain
(around $\theta=-0.5$) and the second, corresponding to partially `convinced'
agents, located around $\theta=0.1$. The decrease of the filter
effectiveness $f$ increases the number of agents in the latter group.
Because the ensemble averaged belief distribution enters the process
for the next iteration, for $f<0.42$, a positive feedback 
mechanism leads to the
eventual dominance of the convinced group. On the other hand, for  $f>0.43$  the size of the
convinced group is too small to persist, and eventually all agents
retain or revert to their leftist stance.

\begin{figure*}
\includegraphics[width=1\textwidth]{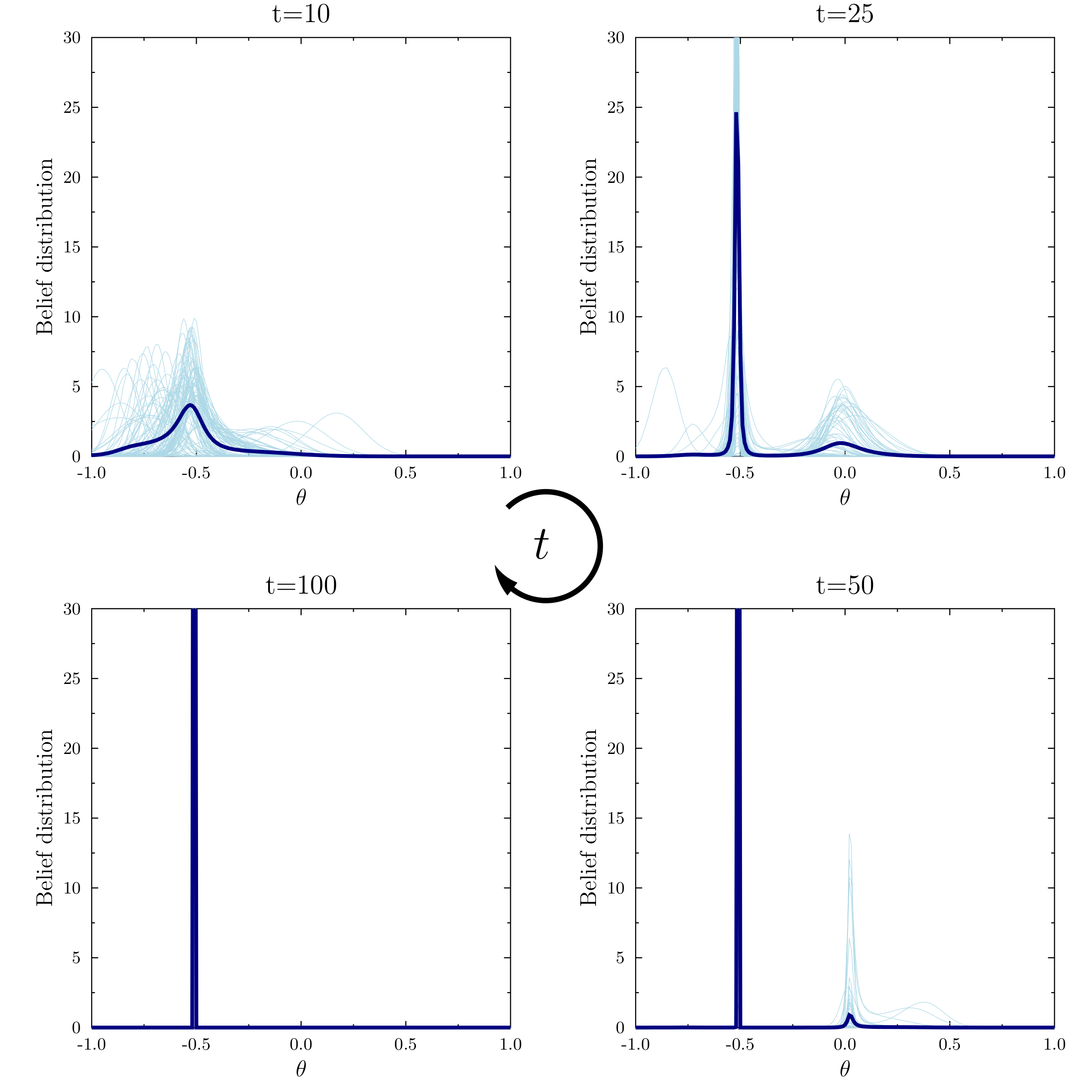}

\caption{\textbf{Politically Motivated Reasoning filtering, perfect memory.} Time snapshots of the individual belief distributions for the PMR
filter for $f=0.43$. The individual agent's belief distributions
at $t\approx25$ are divided into the `inflexibles' \textendash{}
with opinions centred around $\theta\approx-0.5$, and the agents who were
influenced by the news source, with their distributions centred around
$\theta\approx0$. Thick dark line shows the average distribution of beliefs
- which serves as a filter for the next time step. For $f$ greater
than 0.43 the number of the influenced agents is too small, and repeated
interactions diminish the influence of the $\theta\approx0$ filter peak.
At $t=100$ all agents revert to the leftist positions. \label{fig:Timef430}}
\end{figure*}

\begin{figure*}
\includegraphics[width=1\textwidth]{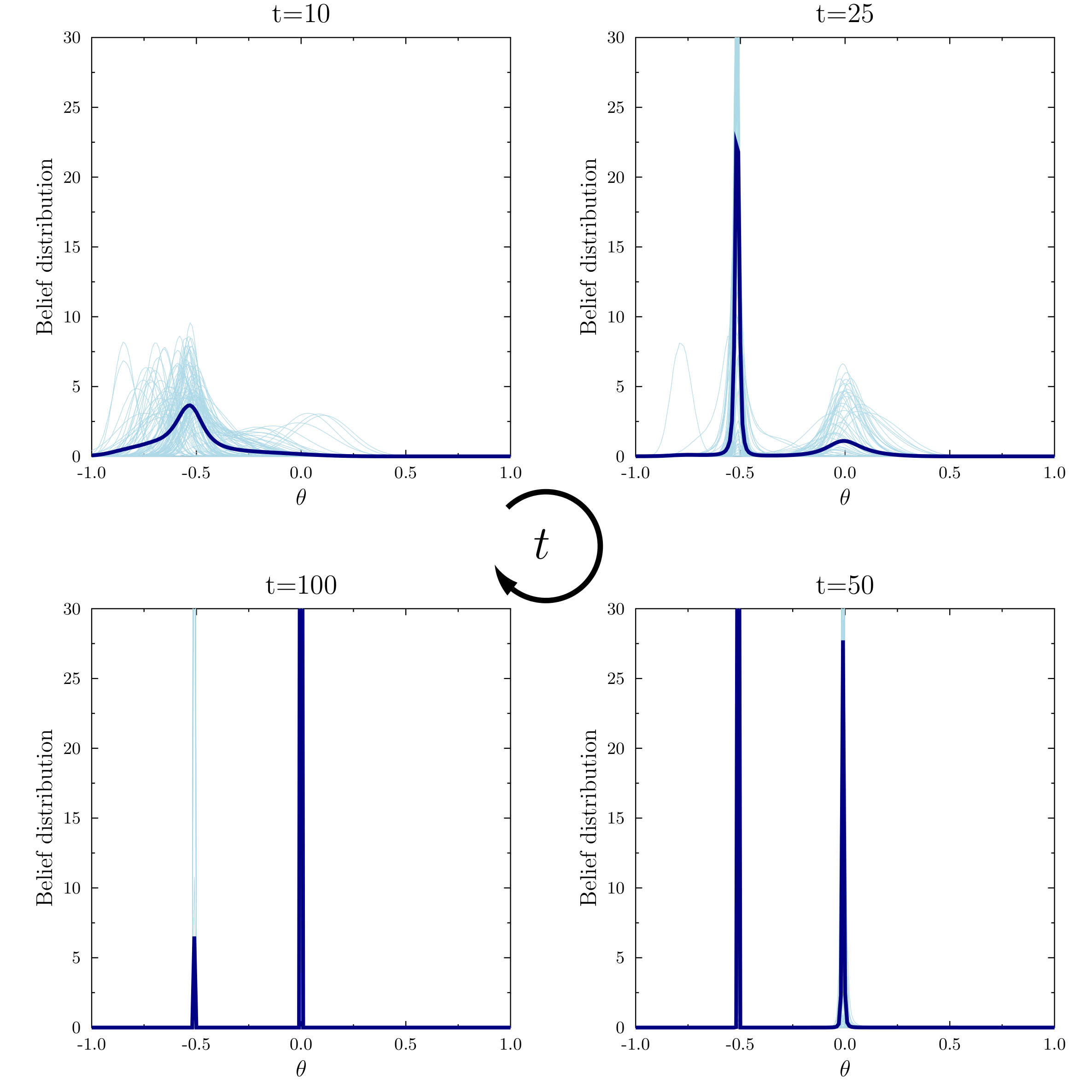}

\caption{\textbf{Politically Motivated Reasoning filtering, perfect memory.} Time snapshots of the individual belief distributions for the PMR
filter for $f=0.42$. As before, the individual agent's belief distributions
at $t\approx25$ are divided into the `inflexibles' \textendash{}
with opinions centred around $\theta\approx-0.5$, and the agents who were
influenced by the news source, with their distributions centred around
$\theta\approx0$. Thick dark line shows the average distribution of beliefs
- which serves as a filter for the next time step. For $f$ smaller
or equal 0.42 the number of the influenced agents becomes large enough
to eventually dominate, and the repeated interactions move all agents
to the centrist position. The jump observed in Figure \ref{fig:case3jump}
occurs when the number of the influenced agents passes the necessary
threshold. Due to the positive feedback, once the peak $\theta\approx0$
dominates the filtering, the repeated filtered information processes
further increase its the size in the subsequent interactions. \label{fig:Timef420}}
\end{figure*}

The results for the Politically Motivated Reasoning filter
were obtained using an assumption that the composition of the group to which an agent
looks for the belief guidance remains unchanged.
The simulations assume that each agent considers the whole group,
defined in the initial input files, to calculate the ensemble averaged
belief distribution $X_{G}(\theta,t)$, which would be used as the filter.
This leads to the case when the more flexible agents, who have shifted
their opinion can eventually pull the whole group with them (for small enough $f$ values). 

Such assumption might be criticized
from a sociological point of view. In a situation, such as that depicted
in $t=50$ panels of Figures \ref{fig:Timef430} and \ref{fig:Timef420},
where the belief systems of the agents flexible and inflexible agents
have very little overlap, one could expect that each of the sub-groups
would \textbf{restrict their PMR filter to the group of the currently
like-minded agents}. In other words, the flexibles, who have moved
away from the initial group average, would be rejected 
by the less flexible agents, as traitors
of the cause, and disregarded when calculating the PMR filter. 
The obvious result would be a split of the initial
group, occurring within just a few filtered iterations (somewhere
between $t=25$ and $t=50$). In such approach it would be useful
to change the simulation measurements from the group averages of belief
$\langle\theta\rangle_{G}$ to the \textbf{numbers} of the 
inflexibles, unconvinced by the information, and the agents who have
shifted their beliefs. Such a dynamical group composition model variant
shall be the topic of later works. 

\subsection{Case 3a: PMR filter with imperfect memory ($m<1$) }

\begin{figure}
\includegraphics[width=1\columnwidth]{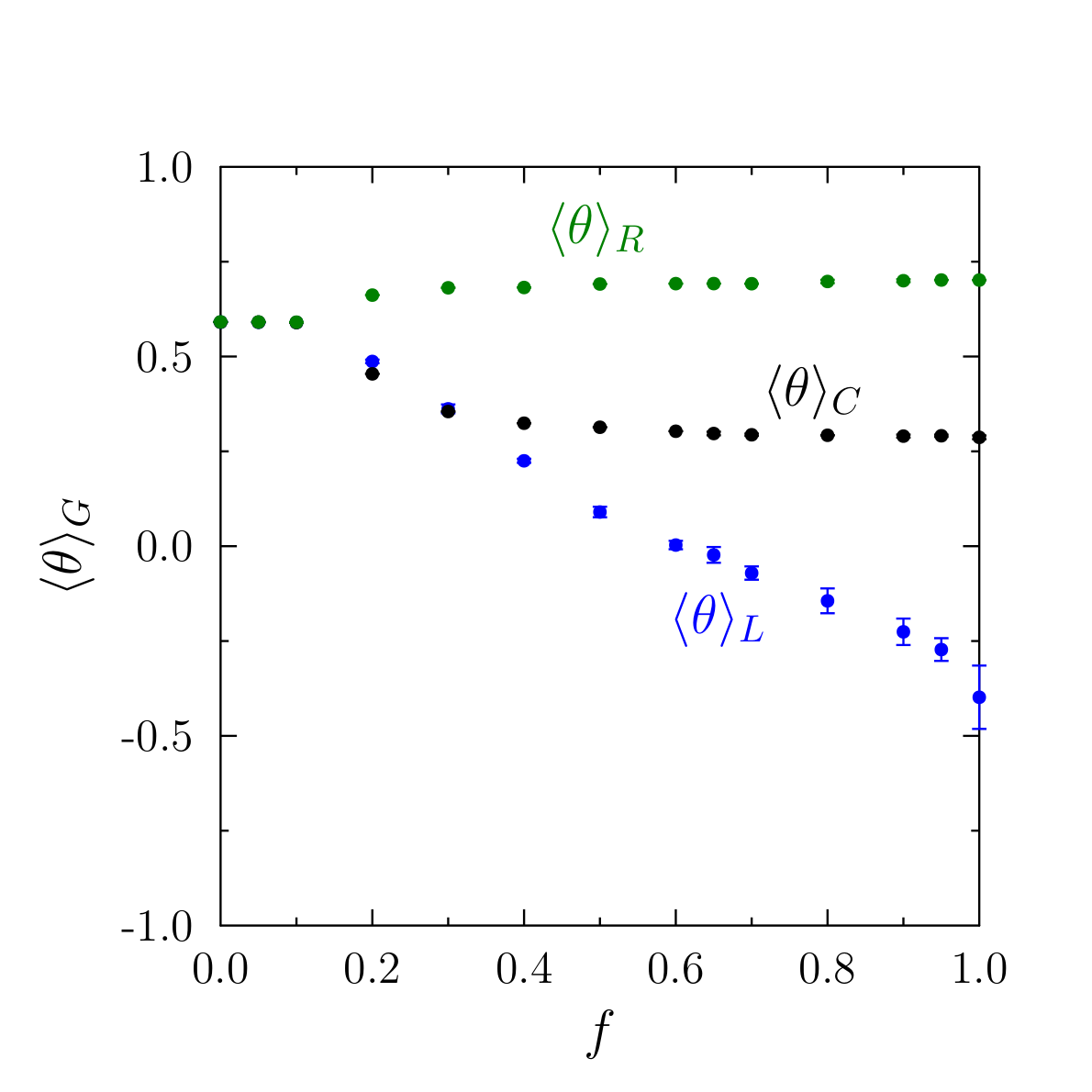}

\caption{\textbf{Politically Motivated Reasoning filtering, reset of belief indeterminacy.} Dependence of the final value of $\langle\theta\rangle_{G}$ for the
three groups, as functions of filtering effectiveness $f$ for the
PMR filter with imperfect memory $m=0.5$. The broadening of the individual
belief distributions due to the imperfect memory restores the almost
linear dependence of the ensemble average value of opinion for the
leftist group. The resulting opinion distribution for the leftist
group $\langle\theta\rangle_{L}$, for $f>0.6$, shows sizeable differences
between the individual simulation runs, which are indicated by  error bars. \label{fig:case3A}}
\end{figure}

The discontinuous change in the system behaviour, described in the previous
section,  results from the extreme narrowing of the
individual belief distributions, due to the repeated application of
the filter. Guided by the analyses of the confirmation
bias filter with imperfect memory, we expect that the reset of individual
belief indeterminacy should significantly change the system behaviour. Figure \ref{fig:case3A},
which presents the results for $m=0.5$, confirms these expectations
are true. Instead of the discrete jump seen for the leftist group
in the unmodified $m=1$ case (Figure \ref{fig:case3jump}), we observe
smooth changes of all group averages of beliefs $\langle\theta\rangle_{G}$.
Moreover,  a full consensus is reached
for finite (although small) values of $f$.
An additional difference in the simulations for the imperfect memory
PMR filter from all cases considered so far, 
is that simulation runs converge
to somewhat different configurations. We
have indicated this as error bars in Figure \ref{fig:case3A}. 

\begin{figure}
\includegraphics[width=1\columnwidth]{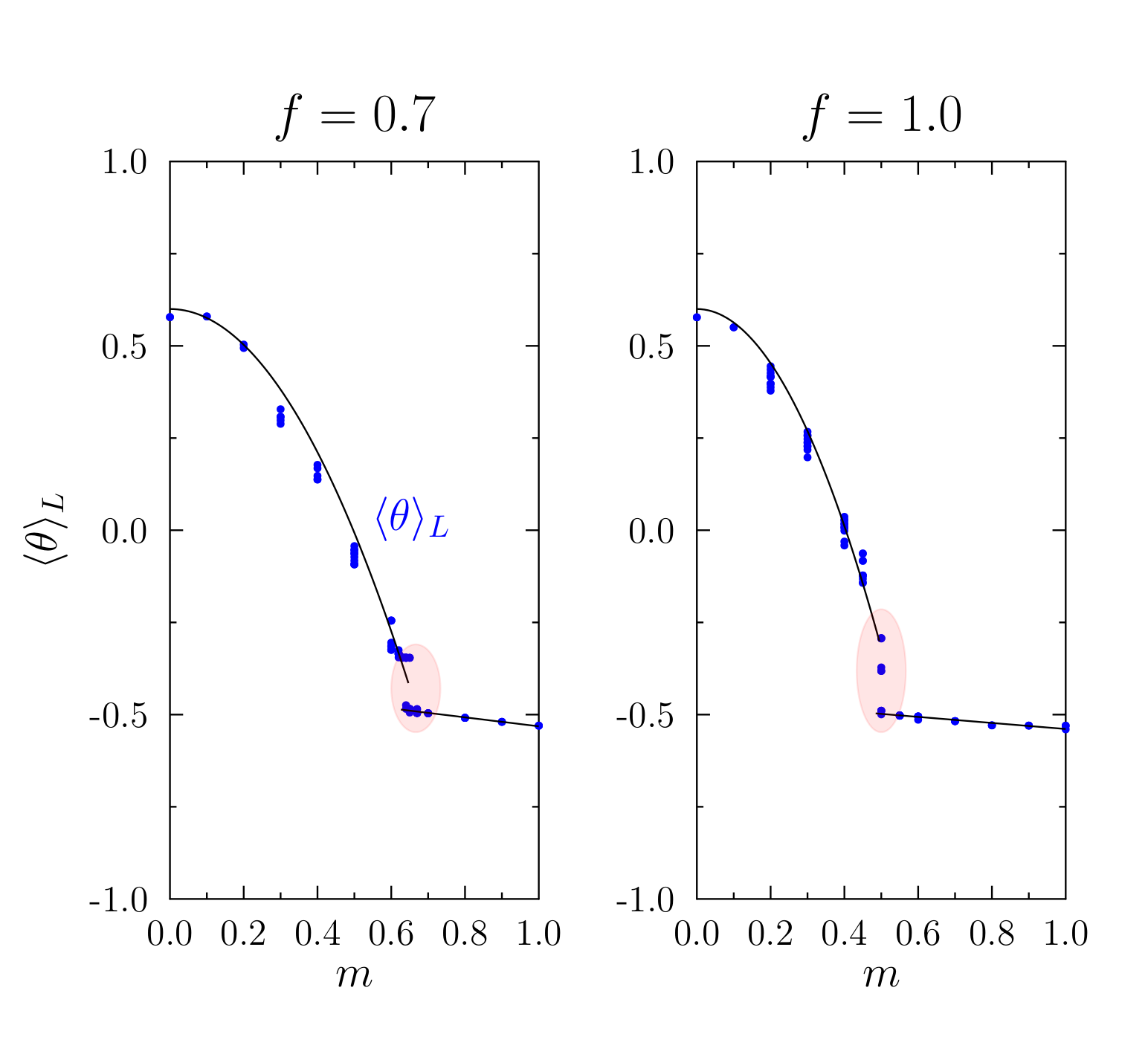}

\caption{\textbf{Politically Motivated Reasoning filtering, memory effects.} Dependence of the final value of $\langle\theta\rangle_{G}$ for the
leftist group on the memory factor $m$, for two values of the filtering
effectiveness $f$ for the PMR filter. Dots show results of the individual
simulation runs. Red ellipses indicate the regions close to the threshold
value of $m$, at which the behaviour of the system changes. Decreasing
the memory quality from the perfect case ($m=1$) leads initially
to very slight, linear shift in the $\langle\theta\rangle_{L}$ value,
attributable to belief changes in the first few interactions. Below
the threshold value (which depends on $f$) the group opinion average
grows to approach the true value of $\theta_{T}$ for $m<0.1$. The
black lines are separate best fits of linear function (for $m$ greater
than a threshold value) and quadratic function for $m$ smaller than
it. \label{fig:case3Am-1} }
\end{figure}

The roughly linear dependence of $\langle\theta\rangle_{L}$ on $f$,
for $f>0.1$, 
results from the increased individual opinion flexibility 
introduced by the admixture of the broad-minded
component of the individual beliefs treated as priors. To better understand
this, we have studied the dependence of $\langle\theta\rangle_{L}$
on the memory factor $m$ for fixed values of $f$. The results are
shown in Figure \ref{fig:case3Am-1}. In the case of relatively effective
PMR filter ($f=0.7$ and $f=1.0$) there are two distinct regimes
of system behaviour. Above certain threshold value $m_T(f)$,  there is only a weak, linear dependence of $\langle\theta\rangle_{L}$
on $m$, mostly due to individual belief shifts during a few initial
time steps, which quickly become frozen. On the other hand, for $m$ smaller than $m_T(f)$, all agents shift their opinions in accordance with the information
source, moving eventually to centrist and rightist positions. The value of $m_T(f)$ is only approximate, as a consequence of the differences
between individual simulation runs, due to the finite size of
the system. 

\section{Discussion }

\subsection{Time dependency considerations}

The choice of the right simulation-to-reality time scaling may depend on
the way we define the information processing events. On one hand,
we could consider only the major news and real world occurrences,
such as the crucial election stories and events. In such a case, the
number of the opinion shaping encounters could be treated as relatively
small, certainly not in the range of tens of thousands or thousands per month.
In such a view, the time periods between the information processing
events are long enough to allow the uncertainty reset. 

At the other end of the spectrum is the vision, in which our beliefs
are shaped by a continuous stream of events, differing in their source
type, intensity, repetition and many other characteristics. Some of
these would originate from external sources, characterized by relatively
stable views and opinions (biased or unbiased at the source), while
other events could originate from more or less random encounters with
other people or observations of ostensibly small importance. In such
microscopic approach, the number of the events could be very large. 

The focus of this work was on the long term effects of a single type
of an information source, interspersed with the periods when the individual
belief structure may become less certain. The goal was to construct
a Bayes based filtered information processing ABM and see if such
approach can yield `reasonable results', by which we mean, depending
on the situation, conditions leading to a general consensus, or, for
other conditions, a persistent disagreement and polarization. The
results have shown that the model can, indeed, produce these results
under simple manipulation of a few key parameters. 

The question of the `right' timescale for opinion change can not be
resolved by such qualitative, simplified model. Among the unknowns are
the effectiveness of the Bayesian update process and the filtering,
the memory imperfection related uncertainty reset scale, and the elements
omitted in the current model, for example differences in the intensity
of particular events. A more realistic model should be based on psychological
studies - which would, hopefully, provide also suggestions as to whether
we should focus on the effects of a few (few tens? hundreds?) information
processing events or to look at the stable or quasi-stable states
reached after thousands of microscopic events. 

\subsection{Manipulation of the Politically Motivated Reasoning Filter }

The current political developments in many democratic societies show
dramatically increasing levels of polarization, covering the general
public and the media (\citet{baldassari07-1,fiorina2008political,bernhardt08-1,stroud10-1,prior2013media,PEW2014political,tewksbury2015polarization}).
In many countries the chances of reaching the state in which a rational
discussions between conflicted groups (not to mention working out
a sensible compromise) seems almost impossible. Recent US presidential
elections provide an obvious example, but the seemingly irrevocable
split exists in many other aspects, sometimes with division lines
not parallel to political ones. A good example of such split is the
existence and (in many countries) growth of the anti-vaccination movements
(\citet{streefland2001public,davies02-1,wolfe02-1,leask06-1,blume2006anti,nelson10-1,kata10-1,betsch11-1,betsch2013debunking,olpinski12-1,mckeever2016silent,telford2016vaccine}),
which are not strictly `politically' aligned. The efforts to convince
vaccination opponents are quite unsuccessful, regardless of the approach
used. Similar problems occur in more politicized issues. This applies
to the cases where suitable evidence is available, for example in
controversies over gun control policies, climate change, GMO, nuclear
energy, and in cases where the beliefs and opinions are largely subjective,
such as evaluations of specific politicians (e.g. Hillary Clinton
or Donald Trump). 

The difficulty in minimizing the polarization may be partially attributed
to the cognitive biases and motivated information processing described
in this paper. Filtering-out of information may be very effective
in keeping a person's beliefs unchanged. In fact, some cognitive heuristics
are evolved to provide this stability (e.g. the confirmation bias).
This makes the task of bridging the gaps between polarized sections
of our societies seem impossible. Still, as Kahan has noted, some
filtering mechanisms may be more flexible than others.

A good example is provided by comparison of the confirmation bias
and PMR. \citet{kahan2016politically1} notes that in some cases PMR
may be confused with the confirmation bias: \emph{Someone who engages
in politically motivated reasoning will predictably form beliefs consistent
with the position that fits her predispositions. Because she will
also selectively credit new information based on its congeniality
to that same position, it will look like she is deriving the likelihood
ratio from her priors. However, the correlation is spurious: a `third
variable'\textemdash her motivation to form beliefs congenial to her
identity \textendash{} is the \textquotedblleft cause\textquotedblright{}
of both her priors and her likelihood ratio assessment. }Kahan notes
the importance of the difference: if the source of the filter is `internal'
(confirmation bias), we have little hope to modify it. On the other
hand, if the motivation for filtering is related to perceptions of
in-group norms, \textbf{the opinions may be changed if the perception
of these in-group norms changes}. Re-framing the issues in a language
that conforms to specific in-group identifying characteristics or
providing information that certain beliefs are `in agreement' with
the value system of the in-group and/or majority of its members, would
change the PMR filtering mechanism. Through this change, more information
could be allowed through, changing the Bayesian likelihood function,
and, eventually, changing the posterior beliefs. 

\subsection{Model extensions and further research directions}

The simulations presented in the current work are based on drastically
simplified assumptions: only a single source of information, with
consistently repeated $S(\theta)$ distribution, only one type of
the filter, our focus is on long term stable conditions. These simplifications
directly indicate the directions of further work: dealing with conflicting
information sources, combinations of different types of filters, transient
phenomena to describe immediate reactions to the exposure of news.
Another planned model extension is related to modelling the possible
dynamical nature of the group norms based PMR filter, mentioned in
Section \ref{subsec:Case-3}. When opinions within a group initially
treated as homogeneous begin to diverge, it is quite likely that the
very definition of the group would change. The agents could redefine
the criteria \emph{who they count as the members of the in-group},
treating those with sufficiently different belief distributions as
outsiders (possibly with a negative emotional label of traitors).
Such a move would dynamically redefine the perceived in-group standards
and norms. The resulting change in the PMR filter could change the
model dynamics from opinion shifts to changes in group sizes and identification. 

The model proposed in this work may be characterised as a `\textbf{reach feature agent}', in contrast to the simplified `spinson' models. 
To examine the possibilities of the approach, we have focused on 
a system in which agents repeatedly react to an unchanging, single external information source. This has allowed to discover some regularities and to understand the roles of the model parameters.

The same general framework of biased processing of information may be used in more complex environments.
It can cover the agents interacting among themselves in arbitrarily chosen social networks.
In such scenario, the input information would be generated by one of the agents (a sender) and would be received and evaluated using the filtering mechanisms and biases by other agent or agents (recipient(s)). Each recipient would then update its opinion (as described by the belief distribution), and, if applicable for the bias type, also the filter function. Of course, it is possible to reverse the roles of the agents and to allow bidirectional communication.
Because the filters used by the communicating agents may be different, the interaction process may be asymmetric. It is also possible to combine the  
agent-to-agent interactions with the influences of external information sources, and to create a truly complex model approximating a real society.

Lastly, especially in the case of the studies of short term, transient
changes, the possibilities of manipulation of the filters by outside
agencies, offer a very interesting and important future research direction.
Such investigations should cover both the manipulations increasing
polarization (partisan information sources and the reliance on emotional
context of the information) as well as the efforts in the opposite
direction -- to detect and to combat the manipulative influences. The latter are especially important to enhance the chances
of a meaningful dialogue in our already highly polarized societies.

\bibliographystyle{plainnat}

\end{document}